\documentclass[aps,pre,twocolumn,groupedaddress,a4letter,showpacs,floatfix]{revtex4}

\usepackage{bm} 
\usepackage{amsmath}
\usepackage{amsfonts}
\usepackage{amssymb} 
\usepackage{longtable}
\usepackage{epsf} 
\usepackage[dvips]{epsfig,color} 
\usepackage{version}
\usepackage{graphicx}
\usepackage{psfrag}

\newcommand{\opencircle}{\mbox{\Large$\circ$}}

\newcommand{\opensquare}{\mbox{$\Box$}}

\newcommand{\opentriangleup}{\mbox{$\triangle$}}
\newcommand{\fulltriangleup}{\mbox{$\blacktriangle$}}
\newcommand{\opentriangledown}{\mbox{$\triangledown$}}

\newcommand{\shortdashedline}{\protect\rule[2pt]{2pt}{1pt}\,\,\protect\rule[2pt]{2pt}{1pt}\,\,\protect\rule[2pt]{2pt}{1pt}}

\newcommand{\dashedline}{\protect\rule[2pt]{4pt}{1pt} \!\protect\rule[2pt]{4pt}{1pt} \!\protect\rule[2pt]{4pt}{1pt}}
\newcommand{\fullline}{\protect\rule[2pt]{15pt}{1pt}}

\newcommand{\rmd}{\mathrm{d}}

\begin{document}
\title{The role of string-like, supramolecular assemblies in reentrant supernematic liquid crystals} 
\author{Marco G. Mazza$^{\dagger}$, Manuel
  Greschek$^{\dagger}$, Rustem Valiullin$^{\ast}$, and Martin Schoen$^{\dagger,\ddagger}$} 

\affiliation{
  $^{\dagger}$Stranski-Laboratorium f\"ur Physikalische und Theoretische Chemie, Technische Universit\"at Berlin, Stra{\ss}e des 17. Juni 135, 10623 Berlin, Germany,\\
  $^{\ast}$Institut f\"ur Experimentelle Physik I, Universit\"at Leipzig, Linn\'estra{\ss}e 5, 04103 Leipzig, Germany,\\
  $^{\ddagger}$Department of Chemical and Biomolecular Engineering, North Carolina State University, 911 Partners Way, Raleigh, NC 27695,
  U.S.A.
  }

\date{\today}
\begin{abstract}
  Using a combination of isothermal-isobaric Monte Carlo and microcanonical
  molecular dynamics we investigate the relation between structure and
  self-diffusion in various phases of a model liquid crystal using the
  Gay-Berne-Kihara potential. These molecules are confined to a mesoscopic
  slit-pore with atomically smooth substrate surfaces. As reported recently
  [see M.~G.~Mazza {\em et al.}, Phys. Rev. Lett. {\bf 105}, 227802 (2010)], a
  reentrant nematic (RN) phase may form at sufficiently high
  pressures/densities. This phase is characterized by a high degree of nematic
  order and a substantially enhanced self-diffusivity in the direction of the
  director $\widehat{\bm{n}}$ which exceeds that of the lower-density nematic
  and an intermittent smectic A phase by about an order of magnitude. Here we
  demonstrate that the unique transport behavior in the RN phase may be linked
  to a confinement-induced packing effect which causes the formation of
  supramolecular, string-like conformations. The strings consist of several
  individual molecules that are capable of travelling in the direction of
  $\widehat{\bm{n}}$ as individual ``trains'' consisting of chains of
  molecular ``cars''. Individual trains run in parallel and may pass each
  other at sufficiently high pressures.
\end{abstract}
\maketitle

\section{Introduction}\label{sec:intro}
A thermodynamic phase is called ``reentrant'' when an order parameter varies
nonmonotonicaly with the thermodynamic field driving the transition.
Reentrancy is ubiquitous in the physics of thermal many-particle systems. It
arises under quite disparate physical conditions encountered, for example, in
quantum gases \cite{kleinert04}, two-dimensional charged colloids
\cite{bechinger00}, or relativistic scalar field models \cite{pinto05}. As far
as soft matter is concerned reentrancy has been reported for self-assembled
supramolecular structures \cite{osaka07,dudowicz09}, wetting phenomena at
oleophilic surfaces \cite{ramos10}, and novel discotic and calamitic liquid
crystals \cite{szydlowska08}. In fact, since the first observation of
reentrant nematic (RN) phases in a seminal paper by Cladis \cite{cladis75}
reentrancy in liquid crystals seems to have received most of the attention.
Liquid crystals offer rich and complex phase diagrams due to the
characteristic interplay of translational and rotational degrees of freedom.
For example, reentrant phase transitions have been reported for the isotropic
phase of mixtures of discotic liquid crystals \cite{lee89}, the ferroelectric
transition in smectic C phases \cite{pociecha01}, the cholesteric-to-blue
phase transition in chiral liquid crystals \cite{heppke90}, and for nematic
(N) phases \cite{sigaud81}.

After the work of Cladis~\cite{cladis75} on binary mixtures, and later in pure
compounds at high pressure and in the supercooled state \cite{cladis77},
Hardouin \emph{et al.}~\cite{hardouin79} showed the occurrence of RN phases
also for pure compounds at atmospheric pressure. For comprehensive reviews the
reader should consult Refs.~\citealp{sigaud81} and \citealp{cladis88}.
Whereas most earlier work on reentrancy of phase transitions in liquid
crystals is experimental in nature comparatively little attention has been
paid to this fascinating phenomenon from a theorist's point of view.  Early
theoretical approaches have mainly considered dimer models and frustration
effects~\cite{longa82,cladis88,berker81,netz92}. Ferrarini {\em et al.}
consider the role of association and isomerization equilibria for the
formation of RN phases within a mean-field-type of theory \cite{ferrarini96}.
The most recent theoretical study employs isothermal-isobaric and canonical
ensemble Monte Carlo (MC) simulations to investigate the nature of the smectic
A(smA)-RN phase transition for a bulk system of hard ellipsoids with
square-well attraction \cite{demiguel05} where earlier theoretical studies are
briefly reviewed, too.  Unfortunately, the model employed in
Ref.~\citealp{demiguel05} is somewhat artificial in assuming that the
ellipsodal molecules are always oriented in a perfectly parallel fashion such
that all rotational degrees of freedom are ``frozen'' irrespective of the
thermodynamic conditions. Therefore, this study seems only of limited use to
elucidate properties of RN phases at a molecular level. Moreover, the authors
do not consider dynamic features of RN phases.

Despite the variety of systems and thermodynamic conditions under which
reentrancy in liquid-crystalline materials arises comparatively little
attention has been paid to the dynamics of reentrant phases. To the best of
our knowledge the only study of the dynamics of reentrant liquid-crystalline
phases has recently been conducted by ourselves and focuses on self-diffusion
in the RN phase as a specific case \cite{mazza10}. In that work we considered
rod-like mesogens where the interaction between a pair of rods takes into
account both a realistic shape of the molecules and the orientation dependence
of their interaction. Unlike de Miguel and Mart{\'i}n del R{\'i}o
\cite{demiguel05} we allowed our molecules to rotate as freely as the specific
thermodynamic conditions permit.  We could then demonstrate \cite{mazza10}
that on account of the high degree of nematic order diffusion of molecules in
the direction of the nematic director may be enhanced by about an order of
magnitude over that characteristic of lower-density N or smA phases. The
combination of a high degree of nematic order with a substantially enhanced
self-diffusivity in the direction of the nematic director prompted us to refer
to liquid crystals in the RN phase as ``supernematics'' \cite{mazza10}.
Enhanced self-diffusivity in the RN phase bears a striking similarity to
``levitation'' in zeolites.  Experiments have demonstrated the self-diffusion
constant to pass through a maximum when the size of a diffusant molecule is
comparable to the pore size, that is enhanced diffusivity is observed as a
result of severe confinement to tiny voids in the host material (see, for
example, Ref.~\citealp{borah10}).

For liquid crystalline materials we could explain this unusually enhanced
self-diffusion quantitatively in terms of a severely reduced rotational
configurational entropy~\cite{mazza10}. The relation between self-diffusion
and structural features of the RN phase has not been explored. In fact, as we
shall show in this work the RN phase is characterized by unique conformations
of entire groups of its molecules to which we refer as ``strings''. Strings
may be thought of as train-like arrangements of several molecules that stay
together as molecular ``cars'' and diffuse as a supramolecular entity. As we
argue below it is the fairly high pressure that keeps the trains together and
which allows them to overcome the attractive interactions between molecules
pertaining to neighboring trains.  From the experimental side there also
exists some evidence suggesting unique dynamic features of the RN phase. For
example, distinct differences in the molecular dynamics in the N and RN phases
can be concluded from corresponding changes in the nuclear magnetic resonance
(NMR) relaxation times reported in
Refs.~\citealp{dong81,dong82,miyajima84,bharatam99}. However, a more detailed
account of the present experimental state of the art is postponed until
Sec.~\ref{sec:sumcon}.

Last but not least we note that an improved understanding of the dynamics of
liquid crystals is not only of academic interest but may also be important
from a purely practical point of view.  For example, by pumping liquid
crystalline material through vesiscles, that maybe perceived as confined
geometries, a spider is capable to spin silk with unsual materials properties
that result essentially from the transport of the liquid crystal.  Materials
with properties comparable to spider silk cannot be produced in the laboratory
to date (see Fig.~2 in Ref.~\citealp{vollrath01}.

The remainder of our manuscript is organized as follows. Our model liquid
crystal is introduced in Sec.~\ref{sec:sim} whereas Sec.~\ref{sec:res} is
given to a presentation of our findings. In Sec.~\ref{sec:sumcon} we summarize
our results, discuss their physical origin, and put them in a broader
experimental-theoretical context.  Specifically, we rationalize the relation
between the formation of supramolecular strings and enhanced self-diffusion in
the RN phase. Details about the theoretical background of our simulations can
be found in Appendix~\ref{sec:theoback}.

\section{Simulational details}\label{sec:sim}
\subsection{Model potentials}\label{sec:mod}
\begin{figure}[htb]
\begin{center}
\epsfig{file=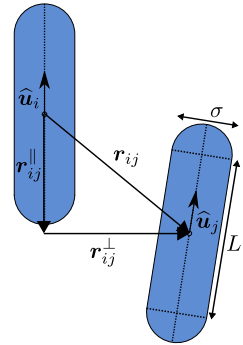,width=0.35\linewidth}
\end{center}
\caption{\small Cartoon of a pair of spherocylinders of length $L+\sigma$
  where $\widehat{\bm{u}}_i$ and $\widehat{\bm{u}}_j$ describe the orientation
  of molecules $i$ and $j$ of that pair in a space-fixed Cartesian coordinate
  system; $\bm{r}_{ij}$ is their center-of-mass distance vector which can be
  decomposed into components (anti-)parallel ($\bm{r}_{ij}^{\parallel}$) and
  perpendicular ($\bm{r}_{ij}^{\perp}$) with $\widehat{\bm{u}}_i$,
  respectively, as indicated.}\label{fig1}
\end{figure}

As in \cite{mazza10} we employ a fluid composed of $N$ spherocylinders of
length $L$ and diameter $\sigma$ capped at both ends by a hemisphere of the
same diameter $2R=\sigma$ (see Fig.~\ref{fig1}). For this system the
Hamiltonian may be cast as
\begin{equation}\label{eq:hamiltonian}
H=
\sum\limits_{i=1}^N
\frac{p_i^2}{2m}+
\sum\limits_{i=1}^N
\sum\limits_{\alpha=x,y}
\frac{\mathcal{L}_{i\alpha}^2}{\mathcal{I}}+
U(\bm{R},\bm{\Gamma})
\end{equation}
where $\bm{P}\equiv(\bm{p}_1,\bm{p}_2,\ldots,\bm{p}_N)$ are the linear momenta
conjugate to the set of center-of-mass positions
$\bm{R}\equiv(\bm{r}_1,\bm{r}_2,\ldots,\bm{r}_N)$ and
$\bm{\Gamma}\equiv(\bm{\gamma}_1,\bm{\gamma}_2,\ldots,\bm{\gamma}_N)$ are the
Euler angles specifying the orientation of each molecule. For the special case
of spherocylinders with two rotational degrees of freedom,
$\bm{\gamma}_i=(\theta_i,\phi_i)$ is a two-dimensional vector composed of two
of the three Euler angles. In Eq.~(\ref{eq:hamiltonian}),
$\mathcal{L}_{i\alpha}$ is the component of the angular momentum referred to
the body-fixed principal axis $\alpha$ of molecule $i$, $m$ is the molecular
mass, and
\begin{equation}\label{eq:momin}
\mathcal{I}=
\frac{3m}{3L+R}
\left(
\frac{8R^3}{15}+\frac{3R^2L}{4}+\frac{RL^2}{3}+\frac{L^3}{12}
\right)
\end{equation}
is the moment of inertia of a spherocylinder.

In a computer simulation [regardless of whether Monte Carlo (MC) or molecular
dynamics (MD) is employed] the key quantity is the configurational potential
energy $U(\bm{R},\widehat{\bm{U}})$ of the system [see, for example,
Eqs.~(\ref{eq:Z}), (\ref{eq:newton2}), or (\ref{eq:torque})] where
$\widehat{\bm{U}}\equiv(\widehat{\bm{u}}_1,\widehat{\bm{u}}_2,\ldots,\widehat{\bm{u}}_N)$
is a set of unit vectors specifying the orientation of the molecules in a
three-dimensional space-fixed Cartesian coordinate system (see
Fig.~\ref{fig1}). Elements of the set $\widehat{\bm{U}}$ are related to
corresponding elements of $\bm{\Gamma}$ via
\begin{equation}
\rmd\widehat{\bm{u}}_i=\sin\theta_i\rmd\theta_i\rmd\phi_i,\quad i=1,\ldots,N
\end{equation}
In the general case of a confined fluid
$U(\bm{R},\widehat{\bm{U}})$ can be decomposed into a fluid-fluid (ff) and a
fluid-substrate (fs) contribution. Assuming pairwise additive interactions the
former can be cast as
\begin{equation}\label{eq:Uff}
U_{\mathrm{ff}}(\bm{R},\widehat{\bm{U}})=
\frac{1}{2}
\sum\limits_{i=1}^{N}
\sum\limits_{j=1\ne i}^{N}
u_{\mathrm{ff}}(d_{ij}^{\mathrm{m}})
\end{equation}
whereas the latter may be expressed as
\begin{equation}\label{eq:Ufs}
U_{\mathrm{fs}}(\bm{R},\widehat{\bm{U}})=
\sum\limits_{k=1}^2
\sum\limits_{i=1}^{N}
u_{\mathrm{fs}}(d_{ik}^{\mathrm{m}})
\end{equation}
In Eqs.~(\ref{eq:Uff}) and (\ref{eq:Ufs}), $u_{\mathrm{ff}}$ and
$u_{\mathrm{fs}}$ represent the specific model potential adopted to describe
the intermolecular interaction between two rodlike molecules and between a
rodlike molecule and a planar, structureless solid substrate, respectively. In
this work we adopt the so-called Gay-Berne-Kihara (GBK) model
\cite{martinez04}. In the GBK model the interaction between a pair of
spherocylinders depends on the relative molecular orientation through the
function
$d_{ij}^{\mathrm{m}}(\bm{r}_{ij},\widehat{\bm{u}}_{i},\widehat{\bm{u}}_{j})$
which depends on the center-of-mass distance vector
$\bm{r}_{ij}\equiv\bm{r}_{i}-\bm{r}_{j}$ and the orientations of molecules $i$
and $j$ such that $d_{ij}^{\mathrm{m}}$ is actually the {\em minimum} distance
between that pair of molecules. More specifically,
\begin{equation}\label{eq:ljmod}
u_{\mathrm{ff}}=
4\varepsilon_{\mathrm{ff}}(\widehat{\bm{r}}_{ij},\widehat{\bm{u}}_{i},\widehat{\bm{u}}_{j})
\left[
\left(\frac{\sigma}{d_{ij}^{\mathrm{m}}}\right)^{12}-
\left(\frac{\sigma}{d_{ij}^{\mathrm{m}}}\right)^{6}
\right]
\end{equation}
where $\widehat{\bm{r}}\equiv\bm{r}/r$ and $r\equiv\left|\bm{r}\right|$. In
Eq.~(\ref{eq:ljmod}) the function
\begin{eqnarray}\label{eq:epsilongb}
\varepsilon_{\mathrm{ff}}(\widehat{\bm{r}}_{ij},\widehat{\bm{u}}_{i},\widehat{\bm{u}}_{j})&=&
\epsilon_{\mathrm{ff}}
\left\{
1-
\frac{\chi^{\prime}}{2}
\left[
\frac{(\widehat{\bm{r}}_{ij}\cdot\widehat{\bm{u}}_{i}+\widehat{\bm{r}}_{ij}\cdot\widehat{\bm{u}}_{j})^2}{1+\chi^{\prime}\widehat{\bm{u}}_{i}\cdot\widehat{\bm{u}}_{j}}+\right.\right.\nonumber\\
&&\left.\left.\frac{(\widehat{\bm{r}}_{ij}\cdot\widehat{\bm{u}}_{i}-\widehat{\bm{r}}_{ij}\cdot\widehat{\bm{u}}_{j})^2}{1-\chi^{\prime}\widehat{\bm{u}}_{i}\cdot\widehat{\bm{u}}_{j}}
\right]
\right\}^2\nonumber\\
&&\times
\frac{1}{\sqrt{1-\chi^2(\widehat{\bm{u}}_{i}\cdot\widehat{\bm{u}}_{j})^2}}
\end{eqnarray}
describes the orientation dependence of the anisotropy strength where the parameters $\chi$ and $\chi^{\prime}$ are given by 
\begin{subequations}\label{eq:chichip}
\begin{eqnarray}
\chi&\equiv&
\frac{\kappa^2-1}{\kappa^2+1}\label{eq:chi}\\
\chi^{\prime}&\equiv&
\frac{\sqrt{\kappa^{\prime}}-1}{\sqrt{\kappa^{\prime}}+1}\label{eq:chip}
\end{eqnarray}
\end{subequations}

\begin{figure}[htb]
\psfrag{U}{$u_{\mathrm{ff}}$}
\psfrag{d}{$d_{ij}^{\mathrm{m}}$}
\begin{center}
\epsfig{file=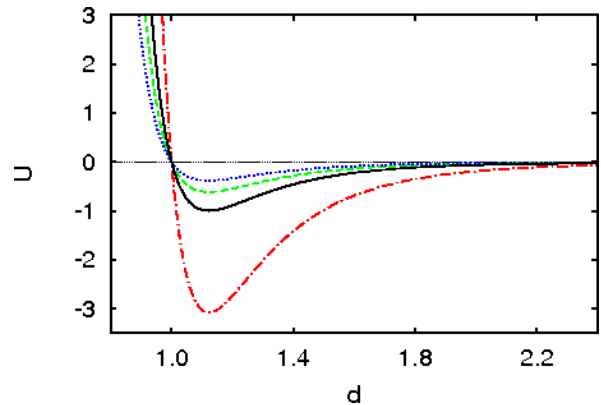,width=0.9\linewidth}
\end{center}
\caption{\small (Color online) Plots of $u_{\mathrm{ff}}$ as functions of
  minimum distance
  $d_{ij}^{\mathrm{m}}(\bm{r}_{ij},\widehat{\bm{u}}_{i},\widehat{\bm{u}}_{j})$
  between a pair of spherocylinders in units of $\epsilon_{\mathrm{ff}}$ and
  $\sigma$, respectively, and for various relative orientations such as
  side-side
  ($\widehat{\bm{r}}_{ij}\cdot\widehat{\bm{u}}_{i}=\widehat{\bm{r}}_{ij}\cdot\widehat{\bm{u}}_{j}=0$,
  $\widehat{\bm{u}}_{i}\cdot\widehat{\bm{u}}_{j}=1$) ($-\cdot-$), end-end
  ($\widehat{\bm{r}}_{ij}\cdot\widehat{\bm{u}}_{i}=\widehat{\bm{r}}_{ij}\cdot\widehat{\bm{u}}_{j}=1$,
  $\widehat{\bm{u}}_{i}\cdot\widehat{\bm{u}}_{j}=1$) (\dashedline), T-shaped
  ($\widehat{\bm{r}}_{ij}\cdot\widehat{\bm{u}}_{i}=0$,
  $\widehat{\bm{r}}_{ij}\cdot\widehat{\bm{u}}_{j}=1$,
  $\widehat{\bm{u}}_{i}\cdot\widehat{\bm{u}}_{j}=0$) ($\cdots$), and crossed
  ($\widehat{\bm{r}}_{ij}\cdot\widehat{\bm{u}}_{i}=0$,
  $\widehat{\bm{r}}_{ij}\cdot\widehat{\bm{u}}_{j}=0$,
  $\widehat{\bm{u}}_{i}\cdot\widehat{\bm{u}}_{j}=0$) (\fullline) (see
  Fig.~\ref{fig1}).}\label{fig2}
\end{figure}

In these latter two expressions parameters $\kappa=L+1$ ($L$ in units of
$\sigma$) and $\kappa^{\prime}$ may be thought of as the aspect ratio of a
spherocylinder and the interaction strength for a side-side relative to an
end-end configuration of a pair of spherocylinders, respectively. This can be
seen from various plots in Fig.~\ref{fig2} indicating that the potential
minimum depends strongly on the molecular orientation such that a side-side
configuration of a molecular pair is the energetically most favored one. One
also notices from Fig.~\ref{fig2} that both the location of the potential
minimum as well as the range of distances over which $u_{\mathrm{ff}}$ becomes
repulsive depend very little on the intermolecular orientation as one would
expect on account of the spherocylindrical shape of the molecules.

In a similar spirit, $d_{ik}^{\mathrm{m}}(z_{ik},\widehat{\bm{u}}_{i})$
introduced in Eq.~(\ref{eq:Ufs}) is the minimum distance between a
spherocylinder with orientation $\widehat{\bm{u}}_{i}$ whose center of mass is
at a vertical distance $z_{ik}=z_i\pm s_{\mathrm{z}}/2$ from the lower ($k=1$)
and upper ($k=2$) substrate located at $-s_{\mathrm{z}}/2$ and
$+s_{\mathrm{z}}/2$ along the $z$-axis of a space-fixed Cartesian coordinate
system, respectively. Assuming the substrates to be planar,
$d_{ik}^{\mathrm{m}}$ can easily be determined as the smaller of the two
distances of both end points of a spherocylinder from the substrate plane. The
corresponding function
$d_{ij}^{\mathrm{m}}(\bm{r}_{ij},\widehat{\bm{u}}_{i},\widehat{\bm{u}}_{j})$
is more complex but can be computed numerically using an efficient algorithm
proposed by Vega and Lago \cite{vega94}. Following \cite{mazza10} we model the
fluid-substrate interaction via
\begin{equation}\label{eq:ufs}
u_{\mathrm{fs}}=
4\epsilon_{\mathrm{fs}}\rho_{\mathrm{s}}
\left[
\left(\frac{\sigma}{d_{ik}^{\mathrm{m}}}\right)^{10}-
\left(\frac{\sigma}{d_{ik}^{\mathrm{m}}}\right)^{4}
g(\widehat{\bm{u}}_i)
\right]
\end{equation}
where the parameter $\epsilon_{\mathrm{fs}}$ controls the strength of
interaction similar to $\epsilon_{\mathrm{ff}}$ in Eq.~(\ref{eq:epsilongb})
and $\rho_{\mathrm{s}}=2/\ell^2$ is the areal density of the substrate where
$\ell/\sigma=1/\sqrt[3]{4}$ is the lattice constant of a single layer of atoms
arranged according to the ($100$) plane of the face-centered cubic lattice.
The diameter $\sigma$ of these substrate atoms is taken to be the same as the
diameter of a spherocylinder of the confined fluid phase. To obtain
Eq.~(\ref{eq:ufs}) we assume the interaction between the point on the
spherocylinder closest to the nearest substrate atom to be described by a
conventional Lennard-Jones potential. Holding both the center-of-mass position
$\bm{r}_i$ and the orientation $\widehat{\bm{u}}_i$ fixed one then averages
the interaction with the substrate atom over the entire substrate area. Notice
that this procedure gives rise to a prefactor $\frac{2}{5}$ in front of the
repulsive term in Eq.~(\ref{eq:ufs}) which we have deliberately omitted for
convenience. 

In Eq.~(\ref{eq:ufs}), $0\le g(\widehat{\bm{u}})\le1$ is the so-called
``anchoring function''.  It permits to discriminate energetically different
orientations of a molecule with respect to the substrate plane. In this paper
we employ
\begin{equation}
g(\widehat{\bm{u}})=
(\widehat{\bm{u}}\cdot\widehat{\bm{e}}_{\mathrm{x}})^2+
(\widehat{\bm{u}}\cdot\widehat{\bm{e}}_{\mathrm{y}})^2
\end{equation}
where $\widehat{\bm{e}}_{\alpha}$ is a unit vector pointing along the
$\alpha$-axis of a space-fixed Cartesian coordinate system. Hence, any
molecular arrangement parallel with the substrate plane is energetically
favored whereas a homeotropic alignment of a molecule
($\widehat{\bm{u}}\parallel\widehat{\bm{e}}_{\mathrm{z}}$) receives an energy
penalty by ``switching off'' the fluid-substrate attraction altogether.

\begin{figure}[htb]
\begin{center}
\epsfig{file=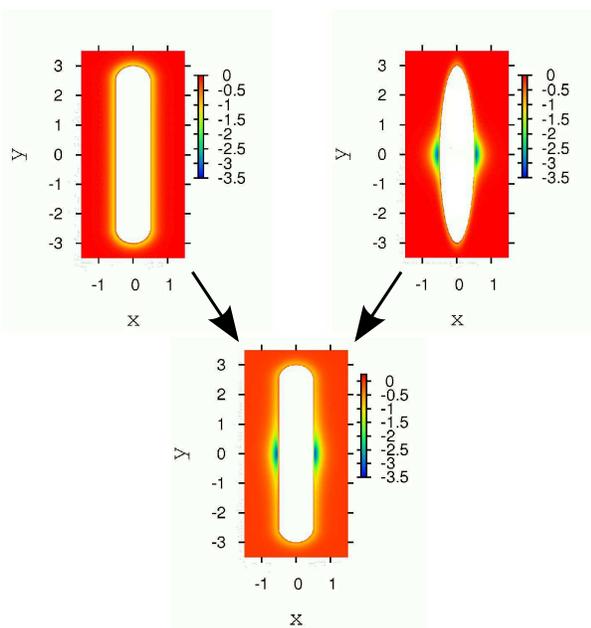,width=1.0\linewidth}
\end{center}
\caption{\small (Color online) Plots of the equipotential contour plot
  $u_{\mathrm{ff}}/\epsilon_{\mathrm{ff}}$ for the GBK (bottom), K (top left),
  and GB (top right) models (see text); color code shown alongside. The area
  shaded in white is defined through the inequaility $u_{\mathrm{ff}}>0$
  therefore approximately representing the molecular shape. In generating the
  plots we assumed a pair of molecules to be located in the $x$--$y$ plane
  with a perfectly parallel orientation (i.e.,
  $\widehat{\bm{u}}_{i}\cdot\widehat{\bm{u}}_{j}=1$) as a special case (see
  also Fig.~\ref{fig1}).  Both $x$ and $y$ are given in units of $\sigma$.
  Note that the GBK model combines the molecular shape of the K model with the
  anisotropy of the GB model.}\label{fig3}
\end{figure}

At this stage a couple of additional comments apply. For example, the reader
may notice that Eqs.~(\ref{eq:epsilongb}) and (\ref{eq:chichip}) are identical
with those describing the orientation dependence of the interaction strength
in the well-known Gay-Berne (GB) model of liquid crystals \cite{gay81}.
However, unlike in the GBK model a liquid crystal molecule is perceived as an
ellipsoid of revolution in the GB model. This is effected by replacing the
function
$d_{ij}^{\mathrm{m}}(\bm{r}_{ij},\widehat{\bm{u}}_{i},\widehat{\bm{u}}_{j})$
in Eq.~(\ref{eq:ljmod}) by one that accounts properly for the shape of a
molecule in the GB model [see, for example, Eq.~(4) of
Ref.~\citealp{gruhn97}]. The GBK model is also closely related to the
so-called Kihara (K) model \cite{kihara63} of a liquid crystal
\cite{cuertos03}. In this latter model a molecule is again perceived as a
spherocylinder but the interaction strength between a pair of molecules
depends only on the minimum distance but not on the relative orientation of
the molecules of a pair. This is achieved by setting $\kappa^{\prime}=1$ in
Eq.~(\ref{eq:chip}) such that
$\varepsilon_{\mathrm{ff}}(\widehat{\bm{r}}_{ij},\widehat{\bm{u}}_{i},\widehat{\bm{u}}_{j})=\epsilon_{\mathrm{ff}}$
in Eq.~(\ref{eq:epsilongb}). The shapes of molecules in the K, GB, and GBK
models resulting from these manipulations are illustrated by the contour plots
in Fig.~\ref{fig3}.

\subsection{Numerical details}\label{sec:numdet}
The simulations (both MC and MD) to be presented below are based upon systems
containing $N=1500$ molecules. Quantities of interest will be expressed in
customary dimensionless (i.e., ``reduced'') units. For example, length will be
expressed in units of $\sigma$, energy in units of $\epsilon_{\mathrm{ff}}$,
temperature in units of $\epsilon_{\mathrm{ff}}/k_{\mathrm{B}}$, time in units
of $(\sigma^2m/\epsilon_{\mathrm{ff}})^{1/2}$ using $m=1$, and pressure
$P_{\parallel}$ in units of $k_{\mathrm{B}}T/\sigma^3$ where 
$P_{\parallel}=\frac{1}{2}(P_{\mathrm{xx}}+P_{\mathrm{yy}})$ is related to
diagonal components of the pressure tensor $\mathbf{P}$ acting in the $x$--$y$
plane.

We employ MC simulations in an isothermal-isobaric ensemble where we use $N$,
$T$, $P_{\parallel}$, and $s_{\mathrm{z}0}$ as input parameters. In all the
simulations we fix $s_{\mathrm{z}0}=19$ and consider $T=4.0$ and $6.0$. To
reduce the computational cost interactions between fluid molecules are cut off
beyond a minimum distance $d_{\mathrm{c}}^{\mathrm{m}}=3$. In addition, we
employ a conventional Verlet neighborlist where molecules up to a minimum
distance of $8$ are included as neighbors \cite{verlet67}. In the $x$- and
$y$-directions periodic boundary conditions are applied.

In MC we employ the algorithm of Schoen \cite{schoen99} with one important
modification. Rather than changing the area $A$ of the simulation box in the
sense of a similarity transformation we allow the individual side lengths
$s_{\alpha}$ ($\alpha=x$, $y$) to vary independently such that the shape of
the simulation cell may change during the course of a simulation. This is
important to preserve the in-plane isotropy of $\mathbf{P}$ especially in
highly ordered N, smA, or RN phases. Our runs start from a sufficiently low
pressure in the isotropic phase. In the initial configuration molecules are
perfectly aligned and then allowed to relax to the isotropic state. The last
configuration of this run is then used as a starting configuration of the next
run at a slightly higher $P_{\parallel}$. We continue with such a compression
sequence until the last state point of the highest pressure of interest has
been reached. We refer to a MC step as $N$ attempted displacements or
rotations and one attempted change of $s_{\alpha}$. Displacements and
rotations as well as changes for $\alpha=x$ or $\alpha=y$ are performed with
equal probability. Our runs are based upon $3\times10^{5}$ equilibration steps
followed by $2\times10^{5}$ production steps for thermodynamic states
sufficiently far away from any phase transition; around these transitions the
number of production steps was enlarged to $7\times10^{5}$. However, in
changing $A$ as described care has to be taken that none of the instantaneous
side lengths $s_{\alpha}$ becomes shorter than twice the neighbor-list cutoff
given above.

MC simulations are mainly employed to generate suitable starting
configurations for the subsequent MD simulations. In addition,
MC is used to independently verify the correctness of the MD
simulations through a comparison of equilibrium properties accessible to both
types of simulations. In MD we employ the velocity Verlet algorithm for linear
molecules in the implementation suggested by Ilnytskyi and Wilson
\cite{ilnytskyi02}. The iterative solution of the equations of motion is
initiated by using the last configuration of the  MC simulation
at the same state point as a starting configuration where velocities are
taken at random from a Maxwell-Boltzmann distribution at the desired
temperature. This starting configuration is equilibrated for another $10^{6}$
time steps of $\delta t=0.02$ using a simple velocity rescaling thermostat.
The equilibrated system is then monitored for another $2\times10^{7}$ time
steps with no applied thermostat. During this part of the run the integration
time step is reduced to $\delta t=10^{-4}$.

\section{Results}\label{sec:res}
\subsection{Structure of ordered liquid-crystalline phases}\label{sec:struc}
To characterize the structure of N, smA, and RN phases we begin by introducing
suitably defined order parameters. For nematic phases this is accomplished via
the so-called alignment tensor defined as \cite{pardowitz80}
\begin{equation}\label{eq:align}
\mathbf{Q}\equiv
\frac{1}{2N}
\sum\limits_{i=1}^{N}
\left(
3\widehat{\bm{u}}_{i}\otimes\widehat{\bm{u}}_{i}-\mathbf{1}
\right)
\end{equation}
where ``$\otimes$'' denotes the direct (i.e., dyadic) product and $\mathbf{1}$
is the unit tensor. Hence, $\mathbf{Q}$ is a real, symmetric, and traceless
second-rank tensor which can be represented by a $3\times3$ matrix. For a
given configuration $(\bm{R},\widehat{\bm{U}})$, $\mathbf{Q}$ can be
diagonalized in the basis of its three eigenvectors $\widehat{\bm{n}}_{-}$,
$\widehat{\bm{n}}_{0}$, and $\widehat{\bm{n}}_{+}$ with associated eigenvalues
$\lambda_{-}$, $\lambda_{0}$, and $\lambda_{+}$. Conventionally
\cite{eppenga84}, one then defines the Maier-Saupe \cite{maier59,maier60}
nematic order parameter $S=\left\langle\lambda_{+}\right\rangle$ where angular
brackets denote an ensemble (MC) or time average (MD). On account of its
definition, $S=0$ in the isotropic phase if the orientation of the molecules
is perfectly random whereas $S=1$ if the molecules are perfectly aligned with
the director $\widehat{\bm{n}}=\widehat{\bm{n}}_{+}$ in the nematic phase.

Even though the nematic phase is characterized by a substantial degree of
orientational order the distribution of center-of-mass positions exhibits the
typical short-range order characteristic of ordinary fluids where the typical
correlation lengths are comparable to the range of intermolecular
interactions. This changes in the smA phase where molecules do not only align
with $\widehat{\bm{n}}$ but develop long-range positional order, that is they
form individual layers in the direction of $\widehat{\bm{n}}$. However,
positional order remains short-range in directions orthogonal to
$\widehat{\bm{n}}$ such that within a given layer the structure remains
fluid-like. Because of the layering of the fluid in the direction of
$\widehat{\bm{n}}$ a suitable quantitative measure of smectic order is given
by the leading coefficient of the Fourier transform of the local density
$\rho(\bm{r}\cdot\widehat{\bm{n}})$
\begin{equation}\label{eq:Lambda}
\Lambda\equiv
\frac{1}{N}
\left\langle
\left|
\sum\limits_{i=1}^{N}
\exp\left[
\frac{2\pi i\left(\bm{r}_{i}\cdot\widehat{\bm{n}}\right)}{d}
\right]
\right|
\right\rangle
\end{equation}
where $d$ is the spacing between adjacent smA layers. If these layers were
ideal, $d=\kappa$ (see Sec.~\ref{sec:mod}). However, in practice
$d\gtrsim\kappa$ on account of thermal fluctuations. Hence, in each
configuration $(\bm{R},\widehat{\bm{U}})$, $d$ is adjusted such as to maximize
$\Lambda$. From Eq.~(\ref{eq:Lambda}) it is also apparent that
$\Lambda\in\left[0,1\right]$. In the nematic phase the superposition of the
complex exponential functions in Eq.~(\ref{eq:Lambda}) causes
$\Lambda\approx0$ because in the direction of $\widehat{\bm{n}}$ a periodic
structural feature of peridodicity $d$ does not exist; by a similar token
$\Lambda\approx1$ in the smA phase.

\begin{figure}
\psfrag{yy}{$S,\Lambda$}
\psfrag{xx}[c]{$P_{\parallel}$}
\begin{center}
\epsfig{file=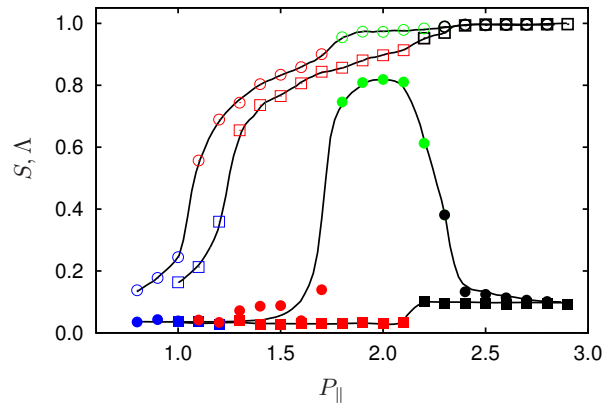,width=0.95\linewidth}
\end{center}
\caption{\small (Color online) Plots of nematic $S$ (open symbols) and smectic
  order parameter $\Lambda$ (filled symbols) as functions of pressure
  $P_{\parallel}$ for various liquid-crystalline phases, $T=4.0$ (circles),
  and $T=6.0$ (squares). I (blue symbols), N (red symbols), smA (green
  symbols), and RN phase (black symbols). The same color code for various
  liquid-crystalline phases is used throughout this work.}\label{fig4}
\end{figure}

Plots of both $S$ and $\Lambda$ are presented in Fig.~\ref{fig4} for
temperatures $T=4.0$ and $6.0$. As one can see from the plots, both $S$ and
$\Lambda$ are small at low pressures indicating the presence of the I phase.
Ideally, $S=0$ in the I phase but in practice attains small positive values of
about $0.2$. This is partly due to the substrates which induce a certain
degree of nematic order in those portions of the confined liquid crystal that
are closest to either substrate. However, a nonvanishing value of $S$ in the
isotropic phase is also due to a small but significant finite-size effect that
is well-known and has been studied quantitatively in the literature
\cite{eppenga84,richter06}. Here we adopt a value $S\simeq0.4$ as a sensible
heuristic definition of the threshold above which the N phase exists in
agreement with the mean-field theory of Maier and Saupe
\cite{maier59,maier60}. From Fig.~\ref{fig4} it therefore appears that the N
phase forms somewhere above $P_{\parallel}\simeq1.0$ ($T=4.0$) and $1.2$
($T=6.0$), respectively.

Focusing on the lower temperature first, $S$ increases in the N phase with
pressure up to $P_{\parallel}\simeq1.7$ indicating the increase of nematic
order. Over the same pressure range there are no smectic layers because
$\Lambda\lesssim0.2$ remains relatively small. However, at
$P_{\parallel}\simeq1.8$, smectic layers are forming as reflected by
$\Lambda\simeq0.8$. Simultaneously, $S$ keeps increasing as well because in
the smA phase molecules align themselves even better with $\widehat{\bm{n}}$.
The smA phase remains stable until $P_{\parallel}\simeq2.2$ has been attained.
Beyond this pressure $S$ remains nearly constant at a high value of about
$1.0$ whereas $\Lambda$ drops rapidly until a residual small value of about
$\Lambda\simeq0.1$ is assumed which remains constant at all higher pressures
considered. The drop in $\Lambda$ with $P_{\parallel}$ clearly reflects the
disappearance of smectic layers present at lower $P_{\parallel}$. Because
$S\simeq1.0$ over the same pressure range we conclude that at sufficiently
high pressures the structure of the confined fluid becomes nematic again in a
reentrant fashion. To distinguish the high-pressure from the low-pressure
nematic we refer to the former as RN because N and RN phases are seperated by
an intermittent smA phase.

At $T=6.0$ these general features of the order-parameter plots in
Fig.~\ref{fig4} prevail except that $\Lambda$ never rises above the residual
value of about $0.1$. Hence, the smA phase does not form at this temperature.
Nevertheless, we notice a small, step-like increase at
$P_{\parallel}\simeq2.2$ in the plot of $\Lambda$. At this and all larger
pressures considered the nematic order is high and increases even further
reflected by a monotonic increase of $S$ toward its limiting value $1.0$. For
$P_{\parallel}\gtrsim2.2$ the confined fluid exhibits structural features of
the RN phase (see below). We shall therefore keep the acronym despite the
missing intermittent smA phase.

\begin{figure}
\begin{center}
\epsfig{file=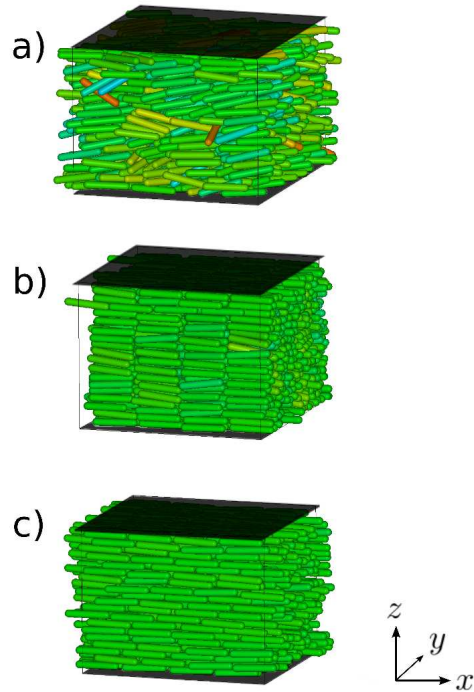,width=0.85\linewidth}
\end{center}
\caption{\small (Color online) ``Snapshots'' of characteristic configurations
  at $T=4.0$; (a) $P_{\parallel}=1.6$ (N), (b) $P_{\parallel}=2.1$ (smA ), (c)
  $P_{\parallel}=2.9$ (RN) (see Fig.~\ref{fig4}). Molecules in green are
  aligned with $\widehat{\bm{n}}$ whereas rectangular areas in dark green
  represent the solid substrates.}\label{fig5}
\end{figure}

To get a better feel for the structure of the confined fluid in the N, smA,
and RN phases we present configuration ``snapshots'' from our simulations in
Fig.~\ref{fig5}. The plot in Fig.~\ref{fig5}(a) illustrates the structure of
the N phase: one clearly recognizes a preferred net orientation of the
molecules but no significant positional order of the center-of-mass
distribution. In Fig.~\ref{fig5}(b) the layer structure characteristic of the
smA phase is visible. These layers disappear in the RN phase as shown in
Fig.~\ref{fig5}(c). However, a large degree of orientational order is
preserved as one realizes from this latter plot.

\begin{figure}
\psfrag{xx}{$z$}
\psfrag{yy}{$\rho\left(z\right)$}
\begin{center}
\epsfig{file=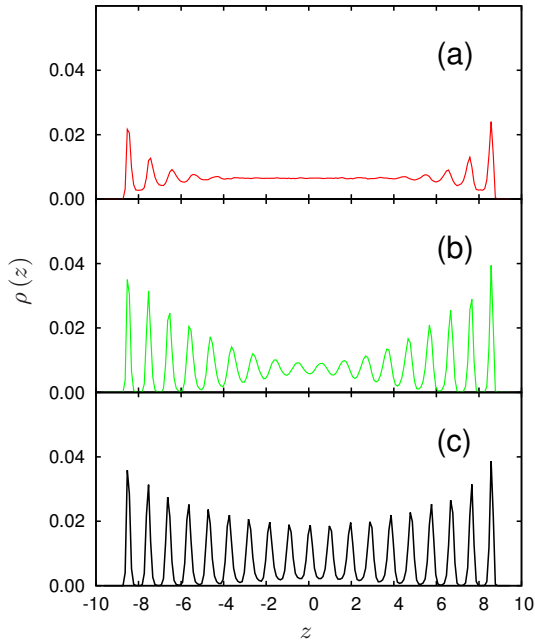,width=0.85\linewidth}
\end{center}
\caption{\small (Color online) Local density $\rho\left(z\right)$ as a
  function of $z$ for $T=4.0$; (a) $P_{\parallel}=1.7$, (b)
  $P_{\parallel}=2.2$, (c) $P_{\parallel}=2.3$ (see Fig.~\ref{fig4} for color
  code).}\label{fig6}
\end{figure}

A more quantitative analysis of the structure of confined liquid crystals in
the various ordered phases is enabled by plots of the local density
\begin{equation}
\rho\left(z\right)=
\left\langle
\sum\limits_{i=1}^{N}
\delta\left(z-z_i\right)
\right\rangle=
\frac{1}{\delta_{\mathrm{z}}}
\left\langle \frac{N\left(z\right)}{A}\right\rangle
\end{equation} 
where $\delta\left(z-z_i\right)$ denotes the Dirac $\delta$-function,
$N\left(z\right)$ is the number of molecules with their centers of mass located
within a interval of width $\delta_{\mathrm{z}}=0.1$, and
$A=s_{\mathrm{x}}s_{\mathrm{y}}$ is the instantaneous area of the simulation
cell of side length $s_{\alpha}$ ($\alpha=\text{x, y}$) in the
$\alpha$-direction. Hence, $\rho\left(z\right)$ is a measure of the
probability of finding the center of mass at a specific position $z$. Plots in
Fig.~\ref{fig6} illustrate the local structure of the confined liquid crystal
in the direction of the substrate normal (i.e., perpendicular to the director
$\widehat{\bm{n}}$). As is well-known from simple fluids \cite{schoen07}, in
confinement $\rho(z)$ is an oscillatory function of position sufficiently
close to the substrate surfaces. The oscillations indicate that molecules
arrange their centers of mass in individual layers parallel with the
substrate. The location of a molecular layer is therefore indicated by a peak
in $\rho\left(z\right)$. In the immediate vicinity of the solid substrate
$\rho\left(z\right)\to0$ because of the diverging fluid-substrate repulsion as
$\left|z\right|\to s_{\mathrm{z}}/2$.

These general features may be seen from the plots in Fig.~\ref{fig6}. One
notices that in the N phase oscillations in $\rho\left(z\right)$ are
restriczed to the immediate vicinity of the solid substrates [see
Fig.~\ref{fig6}(a)]. The oscillations decay exponentially as
$\left|z\right|\to0$ with a correlation length determined by bulk properties
\cite{klapp08} whereas the spacing between neighboring layers reflects the
molecular shape \cite{greschek10}. In the N phase a bulk-like region of
considerable width exists in which $\rho\left(z\right)$ is constant as the
plot in Fig.~\ref{fig6}(a) reveals. On the contrary, layering is much more
pronounced in the smA phase as a comparison of plots in Figs.~\ref{fig6}(a)
and \ref{fig6}(b) indicates: Peaks in Fig.~\ref{fig6}(b) are taller and because
a bulk-like region is absent the entire confined fluid is organized in layers.
However, the layered structure becomes weaker as one approaches the center of
the slit pore as reflected by the decay of $\rho\left(z\right)$ in
Fig.~\ref{fig6}(b). Finally, in the RN phase layering is even more pronounced
as the plot in Fig.~\ref{fig6}(c) shows. Moreover, plots in Fig.~\ref{fig6}(b)
and \ref{fig6}(c) reveal another subtle structural change. Counting maxima in
the plots of $\rho\left(z\right)$ one realizes that one additional layer of
molecules is present in the RN compared with the smA phase. The presence of
this additional layer points to a more efficient packing of molecules as far
as the arrangement of their centers of mass is concerned. A more efficient
packing becomes possible because the layered structure characteristic of the
smA phase [see Fig.~\ref{fig5}(b)] is lost in the RN phase [see
Fig.~\ref{fig5}(c)].

\begin{figure}[htb]
\psfrag{xx}{$z$}
\psfrag{yy}{$\rho\left(z\right)$}
\begin{center}
\epsfig{file=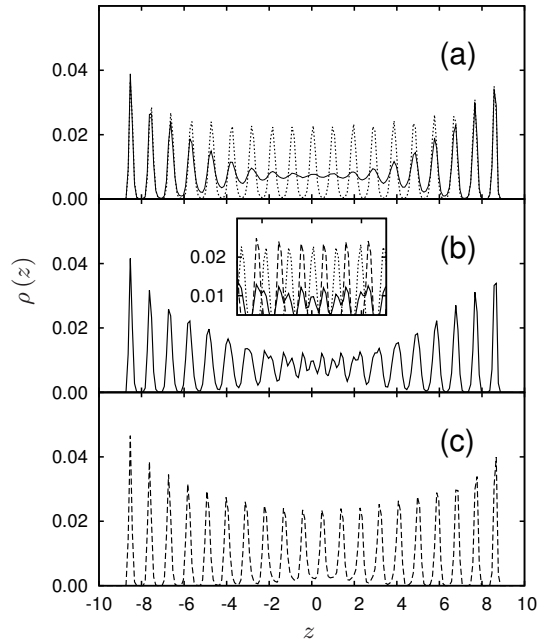,width=0.85\linewidth}
\end{center}
\caption{\small As Fig.~\ref{fig6}, but for $T=6.0$; (a) $P_{\parallel}=2.3$
  (\fullline) and $2.4$ (\shortdashedline), (b) $P_{\parallel}=2.5$, (c)
  $P_{\parallel}=2.6$ (see Fig.~\ref{fig4} for color code). Note that all
  parts of the figure refer to RN states. Inset is a magnification of curves
  for $P_{\parallel}=2.4$, $2.5$, and $2.6$ around $z=0$.}\label{fig7}
\end{figure}

Next, at $T=6$ one realizes from Fig.~\ref{fig7} a similar formation of an
additional molecular layer but without occurrence of an intermittent smectic
phase. Plots in Figs.~\ref{fig7}(a) show that for two state points in the
nematic phase an increase in pressure only causes the layers already present
to become more distinct. For example, $\rho\left(z\right)$ drops to zero in
regions between individual layers at the higher pressure whereas
$\rho\left(z\right)$ remains nonzero over the corresponding ranges at the
lower pressure. The plot in Fig.~\ref{fig7}(b) corresponds to a state point
where the confined fluid undergoes the subtle reorganization that will
eventually lead to the growth of an additional layer. As one approaches the
midplane located at $z=0$ individual peaks in $\rho\left(z\right)$ first
broaden and eventually split up into two unresolved peaks. At an even higher
pressure considered in Fig.~\ref{fig7}(c) we realize that the fluid has
completed its structural reorganization and a new fluid layer emerges as we
can see by comparing this plot with its counterparts shown in
Fig.~\ref{fig7}(a). Moreover, visual inspection of a characteristic
configuration [see Fig.~\ref{fig5}(c)] reveals that in the RN phase molecules
appear in string-like arrangements.
 
\begin{figure}
\begin{center}
\epsfig{file=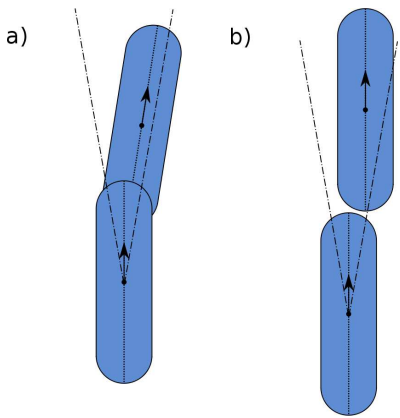,width=0.6\linewidth}
\end{center}
\caption{\small (Color online) Sketch of a pair of spherocylinders (see
  Fig.~\ref{fig1}). (a) Shifted side-side conformation, (b) As (a), but for a
  string-like conformation. The wedge-shaped areas corresponds to the solid
  angle $\widehat{\bm{u}}_i\cdot\widehat{\bm{u}}_j\ge\delta_{\cos\theta}$ used
  in Eq.~(\ref{eq:tilderperp}) which we exaggerate for illustrative
  purposes.}\label{fig8}
\end{figure}

In our terminology a ``string'' is a sequence of molecules arranged in space
like individual cars of a train with their long axes all pointing in the same
direction. The ``snapshot'' of a typical configuration presented in
Fig.~\ref{fig5}(c) already indicates that these supramolecular strings are
omnipresent in the RN phase. In mathematical terms an ideal string can
therefore be defined through the relations
\begin{subequations}\label{eq:uuur}
\begin{eqnarray}
\widehat{\bm{u}}_i\cdot\widehat{\bm{u}}_j&=&1\label{eq:uu}\\
\widehat{\bm{u}}_i\cdot\widehat{\bm{r}}_{ij}&=&
\widehat{\bm{u}}_j\cdot\widehat{\bm{r}}_{ij}=1\label{eq:ur}
\end{eqnarray}
\end{subequations}
where, of course, one of the two scalar products in Eq.~(\ref{eq:ur}) is
redundant. In practice, however, small deviations from perfect alignment of
molecules in a string arise on account of thermal fluctuations. A good measure
of the extent to which a pair of molecules is arranged in a string-like
conformation {\em approximately} may be defined by decomposing the distance
vector $\bm{r}_{ij}$ between the centers of mass of nearest-neighbor molecules
$i$ and $j$ into components parallel and perpendicular to $\widehat{\bm{u}}_i$
through the expressions (see Fig.~\ref{fig1})
\begin{subequations}\label{eq:paraperp}
\begin{eqnarray}
\bm{r}_{ij}^{\parallel}&=&\widehat{\bm{u}}_i\cdot\bm{r}_{ij}\label{eq:para}\\
\bm{r}_{ij}^{\perp}&=&\bm{r}_{ij}-(\widehat{\bm{u}}_i\cdot\bm{r}_{ij})\widehat{\bm{u}}_i\label{eq:perp}
\end{eqnarray}
\end{subequations} 
For a string-like configuration of a pair of molecules one anticipates
$r_{ij}^{\perp}\equiv\left|\bm{r}_{ij}^{\perp}\right|\to0$. However, a second
characteristic conformation characterized by relatively small values of
$r_{ij}^{\perp}$ is conceivable. In this conformation molecules are arranged
in a {\em shifted} side-side arrangement [see Fig.~\ref{fig8}(a)]. Plots of
configurational snapshots in Fig.~\ref{fig5} clearly show that also {\em
  unshifted} side-side arrangements exist both in the N and in the RN phase.
This is because they are energetically favored according to
Fig.~\ref{fig3}(a). However, they are also present to some extent in other
phases as snapshots in Figs.~\ref{fig5}(a) and \ref{fig5}(c) indicate. To
eliminate these unwanted unshifted side-side conformations in favor of only
shifted ones it is prudent from an operational point of view to consider only
the subset
$\left\{\widetilde{\bm{r}}_{ij}^{\perp}\right\}\subseteq\left\{\bm{r}_{ij}^{\perp}\right\}$
defined through
\begin{eqnarray}\label{eq:tilderperp}
\left\{\widetilde{\bm{r}}_{ij}^{\perp}\right\}&\equiv&
\{
\bm{r}_{ij}
|
r_{ij}\le r_{\mathrm{c}}^{\perp}=\kappa+\delta_{\kappa}\wedge
r_{ij}^{\parallel}\ge\frac{\kappa}{2}+1\nonumber\\
&&\wedge\,
\widehat{\bm{u}}_i\cdot\widehat{\bm{u}}_j\ge\delta_{\cos\theta}
\}
\end{eqnarray}
where we arbitrarily take $\delta_{\kappa}=0.2$ and
$\delta_{\cos\theta}=0.98$. To see that this rather complex definition selects
only those conformations sketched in Figs.~\ref{fig8}(a) and \ref{fig8}(b) one
needs to realize that the first of the three conditions in
Eq.~(\ref{eq:tilderperp}) selects nearest neighbors $\{j\}$ of a given
molecule $i$, the second condition favors head-tail-like conformations,
whereas the third one selects molecules with nearly perfectly aligned long
molecular axes. It is then convenient to introduce the probability
\begin{equation}\label{eq:prob}
\mathcal{P}(r^{\perp})=
\frac{\mathcal{P}_0}{N}
\left\langle
\sum\limits_{i=1}^{N}
\sum\limits_{j=1\ne i}^{N}
\delta(r^{\perp}-\widetilde{r}_{ij}^{\perp})
\right\rangle
\end{equation}
for a quantitative discussion of the formation of string-like structures where $\mathcal{P}_0$ is a normalization constant which we determine via
\begin{equation}
\int\limits_{0}^{r_{\mathrm{c}}^{\perp}}\rmd r^{\perp}\,
\mathcal{P}(r^{\perp})
\stackrel{!}{=}1
\end{equation}
In the actual simulations we obtain $\mathcal{P}(r^{\perp})$ as a histogram using $\delta r^{\perp}=0.02$ as the bin width.

\begin{figure}
\psfrag{yy}{$\mathcal{P}(r^{\perp})$}
\psfrag{xx}{$r^{\perp}$}
\begin{center}
\epsfig{file=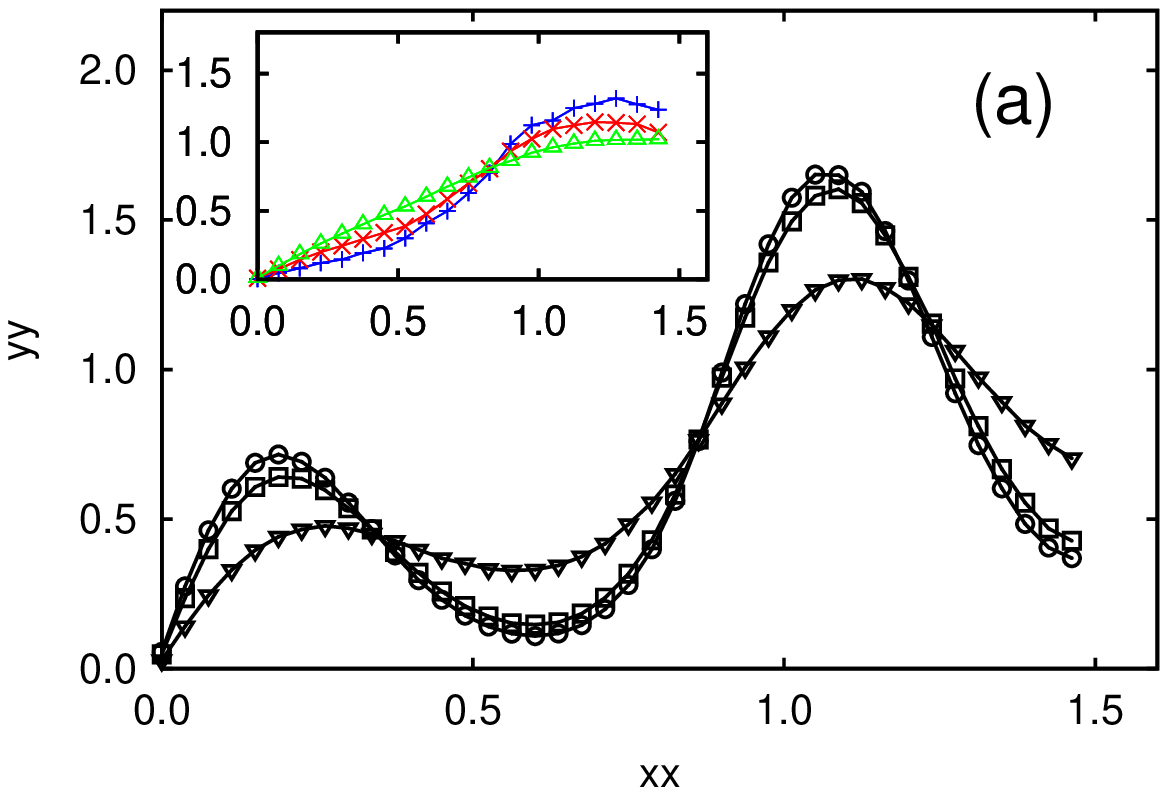,width=0.95\linewidth}
\epsfig{file=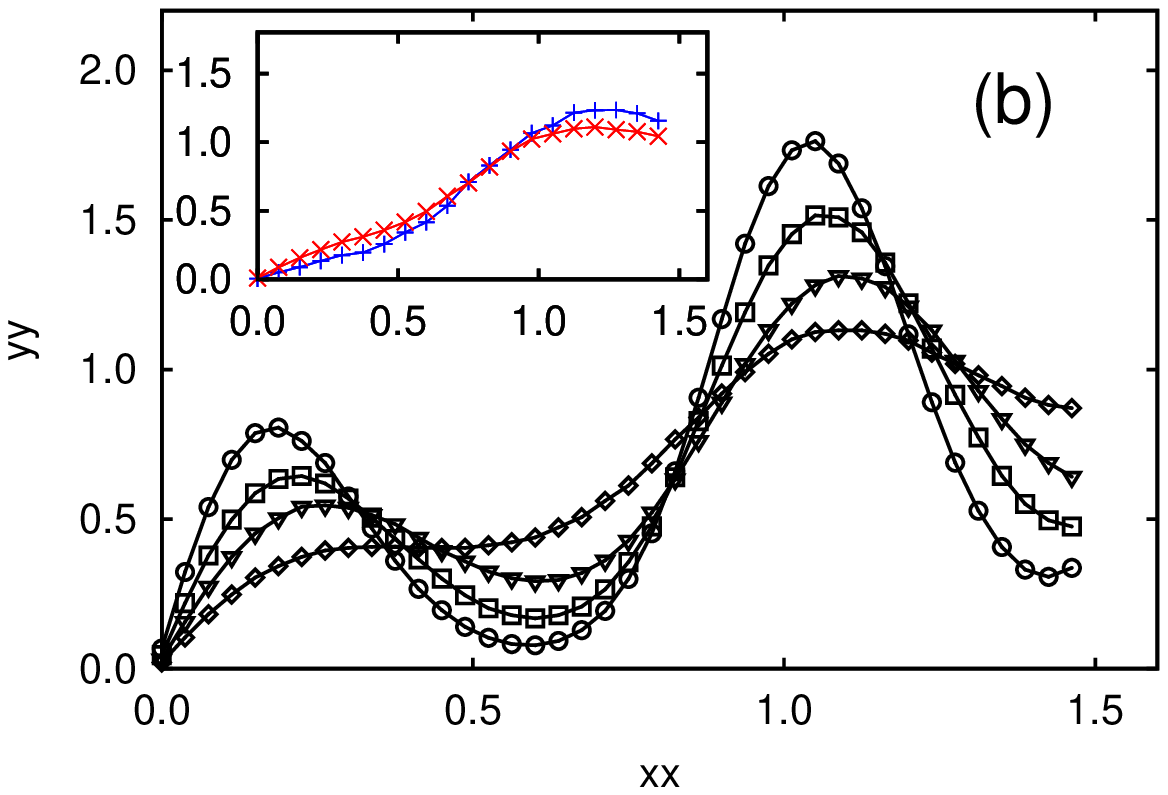,width=0.95\linewidth} 
\end{center}
\caption{\small (Color online) Probability distribution $\mathcal{P}$ of
  $r^{\perp}$ (see Figs.~\ref{fig1} and~\ref{fig8},
  Eqs.~(\ref{eq:paraperp})--(\ref{eq:prob})]. (a) $T=4.0$ and
  $P_{\parallel}=2.3$ (\opentriangledown),  $P_{\parallel}=2.6$ (\opensquare),
  and $P_{\parallel}=2.8$ (\opencircle); inset: $P_{\parallel}=0.9$ ($+$),
  $P_{\parallel}=1.4$ ($\times$), and $P_{\parallel}=1.9$
  (\opentriangleup). (b) $T=6.0$ and $P_{\parallel}=2.2$ ($\diamond$),
  $P_{\parallel}=2.4$ (\opentriangledown), $P_{\parallel}=2.5$ (\opensquare),
  and $P_{\parallel}=2.9$ (\opencircle); inset: $P_{\parallel}=1.1$ ($+$) and
  $P_{\parallel}=1.7$ ($\times$). For line and symbol color code see
  Fig.~\ref{fig4}.}\label{fig9}
\end{figure}

Plots in Fig.~\ref{fig9} show $\mathcal{P}(r^{\perp})$ for different
temperatures and pressures in the I, N, smA, and RN phase. At sufficiently low
pressures it is evident from the insets in Figs.~\ref{fig9}(a) and
\ref{fig9}(b) that $\mathcal{P}(r^{\perp})$ increases monotonically with
$r^{\perp}$ and reaches a broad maximum located approximately at
$r^{\perp}\simeq1.25$ irrespective of whether the I, N, or smA phase is
considered. That $\mathcal{P}(r^{\perp})$ rises with $r^{\perp}$ may be
ascribed to molecules in the I, N, and smA phase that arrange their long axes
locally to a certain degree. This local order is solely caused by the
intermolecular interaction potential which favors local side-side
conformations of molecules [see Fig.~\ref{fig3}(a)].

In the RN phase the form of $\mathcal{P}(r^{\perp})$ is completely different.
Here the plots in Fig.~\ref{fig9}(a) and \ref{fig9}(b) exhibit two maxima. The
largest one slightly above $r^{\perp}\simeq1.0$ corresponds to conformations
in which neighboring particles are arranged in the shifted side-side
conformation depicted schematically in Fig.~\ref{fig8}(a). Note that these
values of $r^{\perp}$ are slightly below the position of the minimum of the
interaction potential (see Fig.~\ref{fig2}). The secondary maximum of
$\mathcal{P}(r^{\perp})$ in the RN phase at $r^{\perp}<0.5$ corresponds to
string-like conformations schematically illustrated by the plot in
Fig.~\ref{fig8}(b). Were the string-like conformation perfect one would
anticipate a peak of $\mathcal{P}(r^{\perp})$ at $r^{\perp}=0$. However, on
account of thermal fluctuations these ideal string-like conformations are only
infrequently observed. For example, closer scrutiny would reveal that
$\mathcal{P}(r^{\perp})\ne0$ at $r^{\perp}=0$ indicating that a very small
fraction of molecules does, in fact, assume an ideal string-like conformation.
As $P_{\parallel}$ increases one notices that both peaks of
$\mathcal{P}(r^{\perp})$ shift to progressively smaller $r^{\perp}$. This fact
reflects an arrangement of molecules in more perfect string-like conformations
with increasing pressure as one would expect intuitively.

One also notices from plots in Figs.~\ref{fig9}(a) and \ref{fig9}(b) that the
curves seem to intersect in isolated points at characteristic values of
$r^{\perp}$. These intersections may be interpreted as ``isosbestic'' points
(derived from the Greek words {\em isos}: equal, the same and {\em sbestos}:
extinguishable) \cite{robinson99}, a concept widely known in spectroscopy. The
occurrence of isosbestic points is relatively simple to grasp. We assume that
$\mathcal{P}(r^{\perp})$ can be decomposed into a sum of three contributions:
\begin{enumerate}
\item{a probability distribution centered at $r^{\perp}\approx1.0$
    corresponding to a population of shifted side-side molecular conformations
    [see Fig.~\ref{fig8}(a)],}\label{it:one}
\item{a probability distribution centered at $r^{\perp}\approx0.25$
    corresponding to a population of shifted string-like molecular
    conformations [see Fig.~\ref{fig8}(b)],}\label{it:two} 
\item{and a probability distribution representing all other molecular
    conformations satisfying Eq.~(\ref{eq:tilderperp}) and embracing the
    entire range of $r^{\perp}$.}\label{it:three} 
\end{enumerate}
For sufficiently large $P_{\parallel}$ the distributions of \ref{it:one} and
\ref{it:two} can be assumed to be well separated whereas \ref{it:three} may be
thought of as a broad background distribution. Intersections between
distributions \ref{it:one} and \ref{it:three} or between \ref{it:two} and
\ref{it:three}, respectively, will give rise to isosbestic points. As
$P_{\parallel}$ increases distributions \ref{it:one} and \ref{it:two} become
sharper and increase in height at the expense of the background distribution
\ref{it:three}; furthermore, maxima of both distributions \ref{it:one} and
\ref{it:two} shift to lower values of $r^{\perp}$ indicating an increase in
order. Importantly, the emergence of two well-defined peaks in
$\mathcal{P}(r^{\perp})$ is a fingerprint of the RN phase and could therefore
be taken as computational evidence to detect RN phases even though its
experimental significance may be limited. Incidentally, at low $P_{\parallel}$
(i.e., in the I, N, or smA phase) the only important distributions are
\ref{it:one} and \ref{it:two} because string-like conformations are
statistically irrelevant. Consequently, only one isosbestic point exists as
shown in the insets of Fig.~\ref{fig9}.

\subsection{Self-diffusivity in liquid crystalline phases}\label{sec:dyn}
\begin{figure}
\psfrag{xx}{$\tau$}
\psfrag{yy}{$\sigma\left(\tau\right)$}
\begin{center}
\epsfig{file=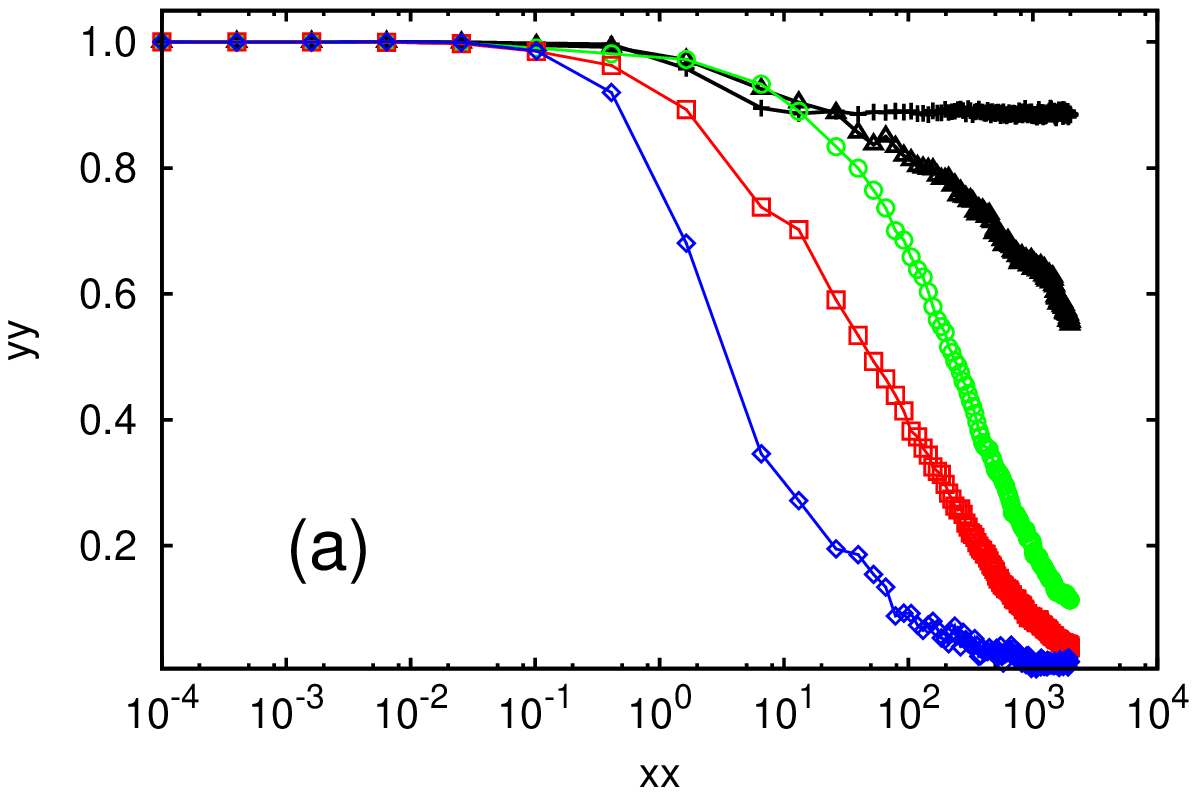,width=0.95\linewidth}
\epsfig{file=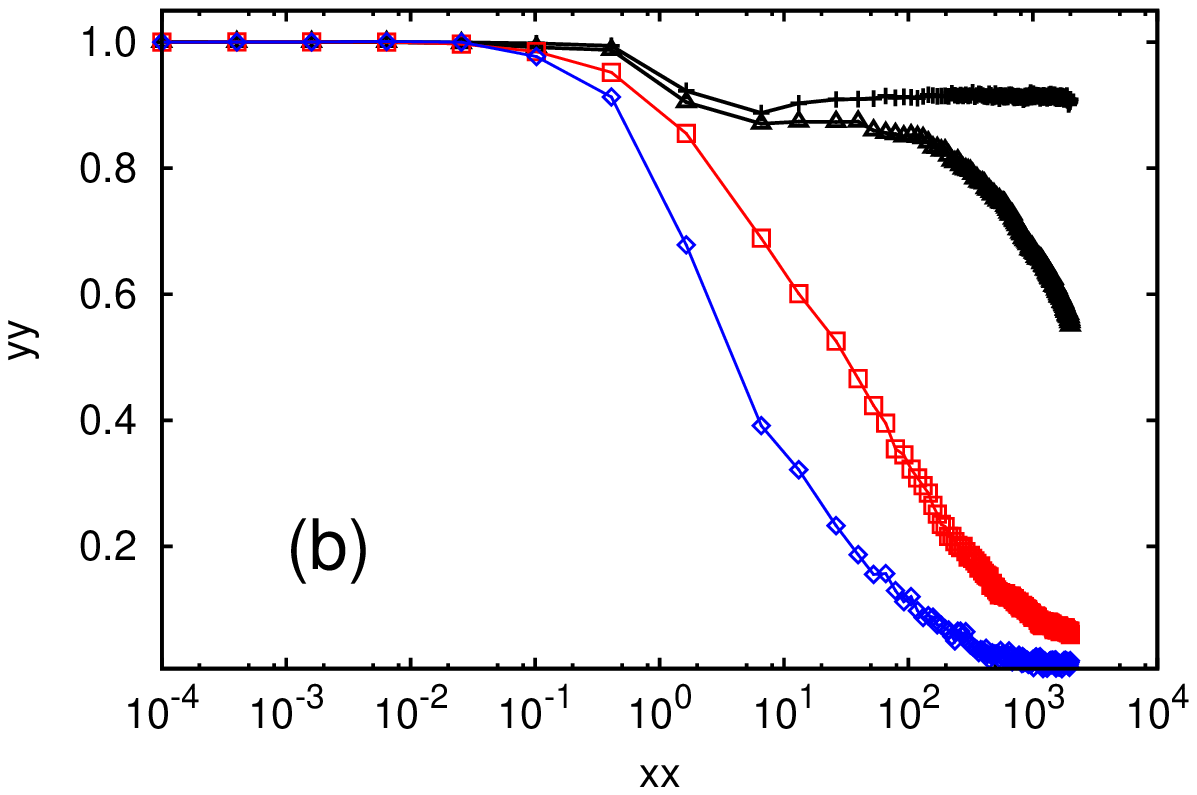,width=0.95\linewidth}
\end{center}
\caption{\small (Color online) (a) Persistence probability
  $\sigma\left(\tau\right)$ as a function of time $\tau$ and $T=4.0$;
  $P_{\parallel}=0.9$ ($\diamond$), $P_{\parallel}=1.7$ (\opensquare),
  $P_{\parallel}=2.1$ (\opencircle), $P_{\parallel}=2.3$) (\fulltriangleup),
  $P_{\parallel}=2.8$ ($+$). (b) as (a), but for $T=6.0$ and
  $P_{\parallel}=1.1$ ($\diamond$), $P_{\parallel}=1.9$ (\opensquare),
  $P_{\parallel}=2.4$ (\fulltriangleup), and $P_{\parallel}=2.9$ ($+$) (see
  Fig.~\ref{fig4} for color code).}\label{fig10}
\end{figure}

The formation of string-like conformations has important ramifications as far as dynamic properties of the RN phase are concerned. To illustrate this we begin by considering a persistence probability 
\begin{equation}\label{eq:persist}
\sigma\left(\tau\right)\equiv
\frac{\displaystyle
\left\langle
\sum\limits_{\left(i,j\right)} 
P_2\left[\eta_{ij}\left(t+\tau\right)\right]
\right\rangle_{t}
}{\displaystyle
\left\langle
\sum\limits_{\left(i,j\right)} 
P_2\left[\eta_{ij}\left(t\right)\right]
\right\rangle_{t}
}
\end{equation}
defined as the correlation function of the angle between a pair of molecules
$(i,j)$ which at the initial time $t$ satisfy the conditions for a string-like
conformation introduced in Eq.~(\ref{eq:tilderperp}). Subscript ``$t$'' is
attached to the angular brackets to emphasize that averaging involves a
sufficiently large number of statistically uncorrelated time origins which
implicitly invokes the principle of stationarity of temporal correlation
functions in equilibrium systems \cite{hansen06}. In Eq.~(\ref{eq:persist})
$\eta_{ij}\equiv \widehat{\bm{u}}_i\cdot\widehat{\bm{r}}_{ij}$ is the cosine
of the angle between the instantaneous orientation of a reference molecule $i$
and the unit vector connecting its center of mass to that of another molecule
$j$ as time passes [see Fig.~\ref{fig1}]. Because of this definition,
$\eta_{ij}\left(t+\tau\right)$ is a measure of the {\em relative} orientation
of a molecular pair at time $t+\tau$ given that its orientation at time $t$ is
$\eta_{ij}\left(t\right)$. Therefore, $\sigma\left(\tau\right)\to1$ as
$\tau\to0$ in agreement with plots in Figs.~\ref{fig10}(a) and \ref{fig10}(b)
for various temperatures and pressures.  At sufficiently low $P_{\parallel}$,
$\sigma\left(\tau\right)$ eventually decays with increasing time. The decay is
delayed the more as the nature of the liquid-crystalline phase changes in the
direction I$\to$N$\to$smA, that is with increasing $P_{\parallel}$ [see
Fig.~\ref{fig10}(a)].

One also notices from Fig.~\ref{fig10}(a) that the decorrelation of relative
orientations becomes very slow as soon as one enters the RN phase. This is
already evident from the plot for $P_{\parallel}=2.3$ immediately after the
smA-RN phase transition has occurred (see Fig.~\ref{fig4}). However, the
corresponding plot in Fig.~\ref{fig10}(a) shows that $\sigma(\tau)$ eventually
decays at long times. This is apparently not so at the higher pressure
$P_{\parallel}=2.8$ in the RN phase. The relevant plot in Fig.~\ref{fig10}(a)
shows only a weak decay of $\sigma\left(\tau\right)$ over an intermediate time
range $1\lesssim\tau\lesssim10$ and then remains nearly constant for longer
times $\tau>10^3$. The shape of $\sigma\left(\tau\right)$ seems to reflect a
two-stage decorrelation process involving those conformations represented by
the bimodal distribution $\mathcal{P}(r^{\perp})$ plotted in
Fig.~\ref{fig9}(a). In fact, intuitively it seems sensible to assume that
decorrelation during the intermediate time interval is associated with the
orientational dynamics of shifted side-side conformations depicted in
Fig.~\ref{fig8}(a). These conformations are expected to be less stable than
the string-like ones depicted in Fig.~\ref{fig8}(b). This notion is further
corroborated by the slowing down of the dynamics at longer times with
increasing $P_{\parallel}$ which causes the population of string-like
conformations to increase according to plots in Figs.~\ref{fig9}. This
qualitative picture remains unaltered at the higher $T=6.0$ as parallel plots
in Fig.~\ref{fig10}(b) show. However, at this temperature the smA phase is
absent in agreement with our earlier analysis of various structural
quantities.

\begin{figure}
\psfrag{xx}[c]{$\tau$}
\psfrag{yy}[c]{$\left\langle\Delta r_{\parallel}^2\left(\tau\right)\right\rangle$}
\begin{center}
\epsfig{file=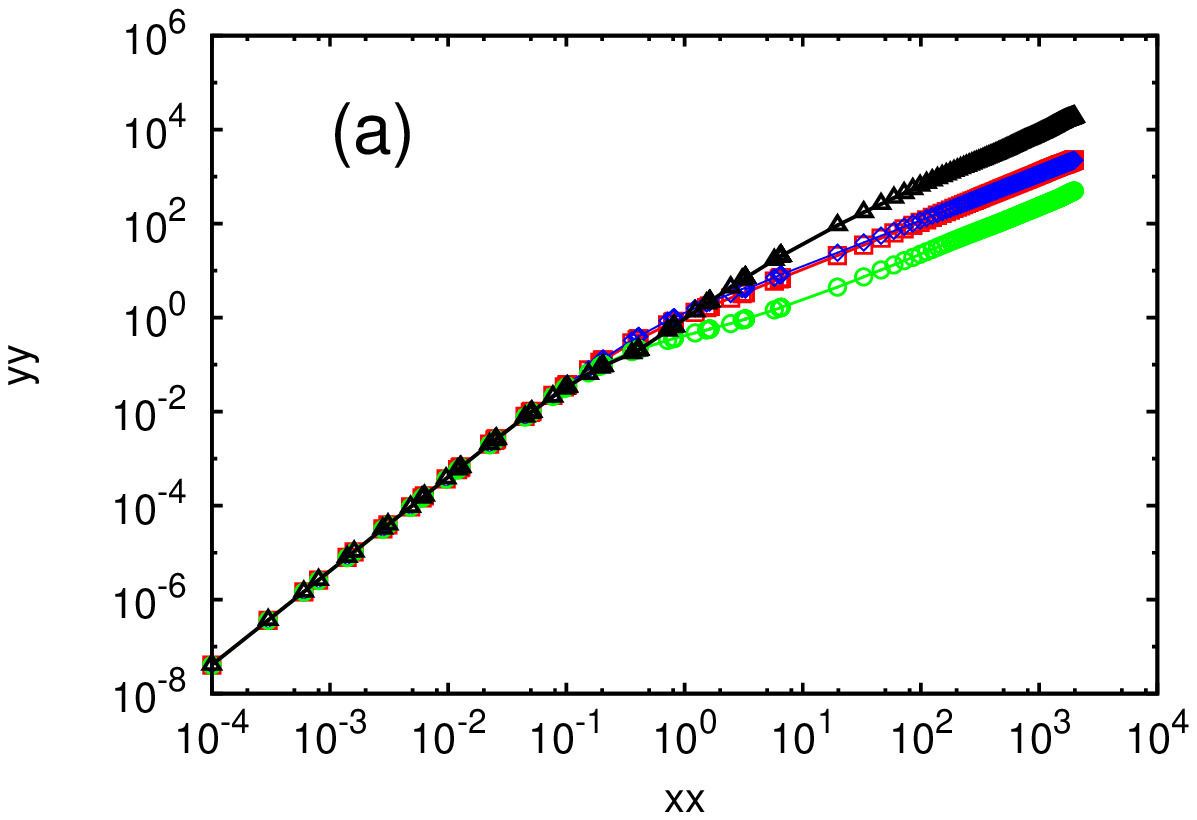,width=0.95\linewidth}
\epsfig{file=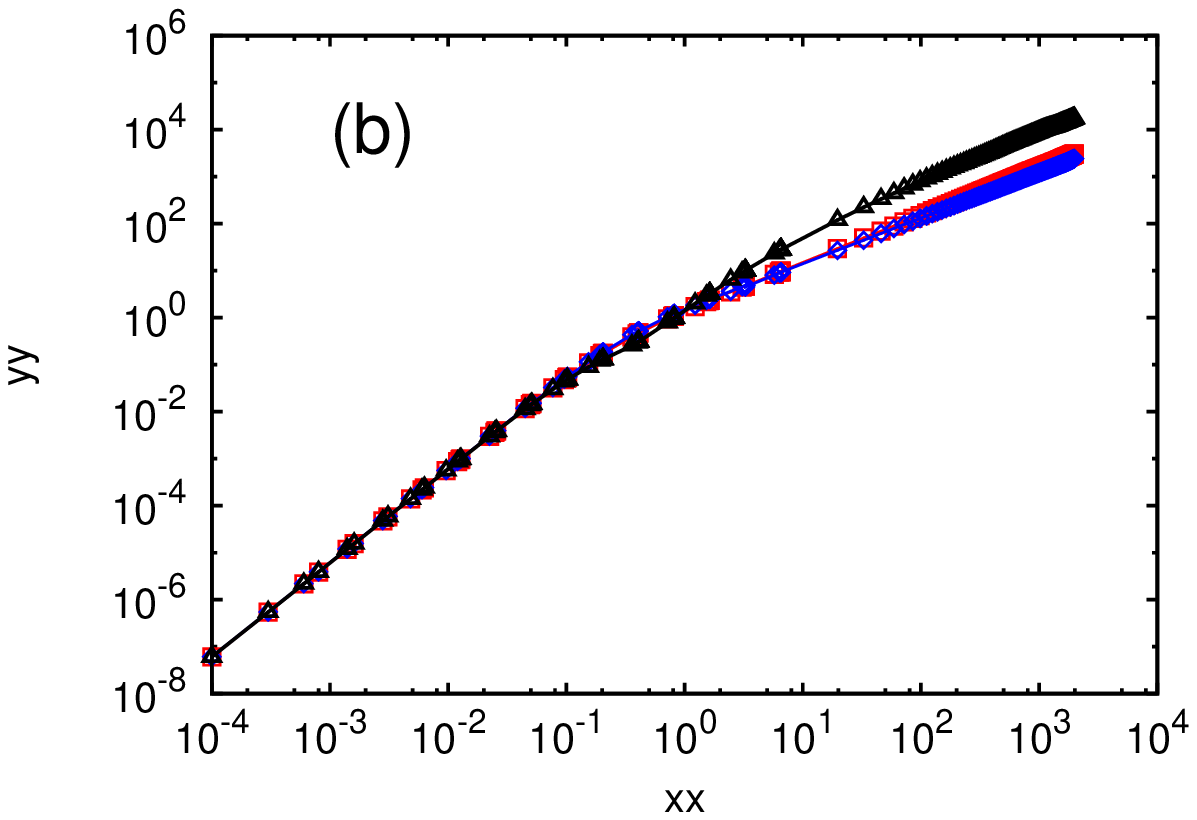,width=0.95\linewidth}
\end{center}
\caption{\small (Color online) (a) MSDs $\left\langle\Delta
    r_{\parallel}^2\left(\tau\right)\right\rangle$ as functions of time $\tau$
  and $T=4.0$ and various liquid-crystalline phases at $P_{\parallel}=0.9$ (I)
  ($\diamond$), $P_{\parallel}=1.7$ (N) (\opensquare), $P_{\parallel}=2.1$
  (smA) (\opencircle), and $P_{\parallel}=2.6$ (RN) (\opentriangleup). (b) as
  (a), but for $T=6.0$ and $P_{\parallel}=1.1$ (I) ($\diamond$),
  $P_{\parallel}=1.9$ (N) (\opensquare), and $P_{\parallel}=2.4$ (RN)
  (\opentriangleup) (see Fig.~\ref{fig4} for color code).}\label{fig11}
\end{figure}

From the discussion of orientational relaxation one may get the impression
that the RN phase is rather sluggish with respect to its dynamics. While this
is certainly true for the rotational dynamics it turns out that mass transport
is actually quite fast in the RN phase. A suitable measure of mass transport
is the mean square displacement (MSD) of molecules. It measures the on-average
net distance a molecule's center of mass is able to travel in a given amount
of time. Because of the large degree of nematic order characteristic of the RN
phase it seems sensible to introduce a {\em specialized} MSD by considering
displacements of molecules in the direction of their long axes. Specifically,
we define $\bm{r}_i^{\parallel}\equiv\widehat{\bm{u}}_i\cdot\bm{r}_i$ such
that the associated MSD for the displacement of molecules in the direction of
their long axes may be cast as
\begin{equation}\label{eq:msd}
\left\langle
\Delta r_{\parallel}^2\left(\tau\right)
\right\rangle_{t}
\equiv
\frac{1}{N}
\left\langle
\sum\limits_{i=1}^{N}
\left[
\bm{r}_i^{\parallel}\left(t+\tau\right)-\bm{r}_i^{\parallel}\left(t\right)
\right]^2
\right\rangle_{t}
\end{equation}
Plots in Fig.~\ref{fig11} illustrate the variation of the MSD with time on a
double-logarithmic scale. At short times $\tau\le10^{-1}$ MSDs for various
phases collapse onto a single curve. In this regime, known as the ballistic
regime, the MSD increases proportionally to $\tau^2$. In the ballistic regime
molecules travel freely, that is for sufficiently short times interactions
with neighboring molecules are inconsequential. As a consequence, the specific
nature of the phase under study does not matter such that the MSDs for
different phases become indistinguishable and can be represented by a unique
curve.

At sufficiently long times, however, the motion of molecules becomes
diffusive, that is $\left\langle\Delta
  r_{\parallel}^2\left(\tau\right)\right\rangle_{t}\propto\tau$. In the
diffusive regime intermolecular interactions do, of course, matter greatly
unlike in the ballistic regime. Therefore, in the diffusive regime plots in
Figs.~\ref{fig11}(a) and \ref{fig11}(b) cannot be represented by a single
curve but differ between I, N, smA, and RN phases. An inspection of the MSD
for the smA phase reveals that ballistic and diffusive regimes are separated
by a third, subdiffusive regime in which $\left\langle\Delta
  r_{\parallel}^2\left(\tau\right)\right\rangle_{t}\propto\tau^{\alpha(\tau)}$
where $1\lesssim\alpha(\tau)\lesssim2$ varies continuously. The crossover
region of subdiffusive behavior reflects the presence of smectic layers
because $\left\langle\Delta r_{\parallel}^2\left(\tau\right)\right\rangle_{t}$
is a measure for mass transport in a direction {\em normal} to the layer plane
[see Fig.~\ref{fig5}(b)]. Diffusive motion of molecules out of their original
smectic layer into a neighboring one is inhibited by the compactness of the
layered structure. However, within each layer there is sufficiently little
order such that molecules may eventually move from their original to a
neighboring layer. This process demarcates the onset of ordinary diffusive
motion. Depending on the local compactness of the smectic layers it may take
some molecules longer than others before they exhibit diffusive motion and
this is the reason why the subdiffusive regime is characterized by a constant
change in the time dependence of the MSD characterized by the exponent
$\alpha(\tau)$.

\begin{figure}
\psfrag{xx}[c]{$P_{\parallel}$}
\psfrag{yy}[c]{$D_{\parallel}$}
\begin{center}
\epsfig{file=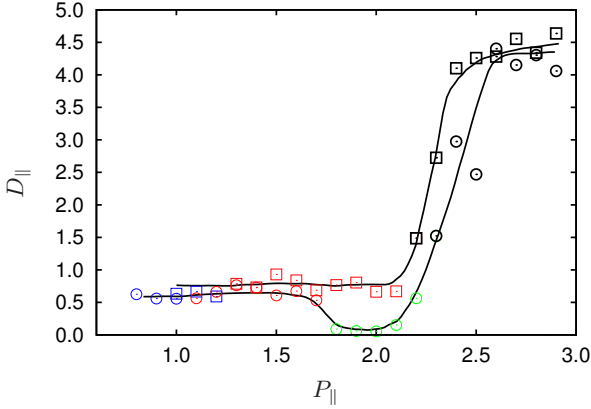,width=0.95\linewidth}
\end{center}
\caption{\small (Color online) Parallel self-diffusion coefficient
  $D_{\parallel}$ as a function of transverse pressure $P_{\parallel}$.
  Circles correspond to $T=4.0$ whereas squares refer to $T=6.0$ (see
  Fig.~\ref{fig4} for color code).}\label{fig12}
\end{figure}

As a quantitative measure of mass transport in the diffusive regime it is convenient to introduce the self-diffusion coefficient $D_{\parallel}$ via the expression
\begin{equation}
D_{\parallel}=
\lim\limits_{\tau\to\infty}
\frac{1}{2\tau}
\left\langle
\Delta r_{\parallel}^2\left(\tau\right)
\right\rangle_{t}
\end{equation}
which corresponds to the long-time slope of the MSDs plotted in
Figs.~\ref{fig11}. From that definition and the plots in Figs.~\ref{fig11} one
immediately anticipates a self-diffusion constant which is largest in the RN
phase compared with the other three phases. In fact, the plot in
Fig.~\ref{fig12} shows a dramatic increase of $D_{\parallel}$ as one enters
the RN phase whereas lower-pressure (I, N, or smA) phases exhibit rather small
self-diffusivity. In particular, the smA phase is characterized by nearly
vanishing self-diffusion constants which can be rationalized as above where we
argued that the relatively compact layered structure makes it difficult for
molecules to diffuse out of their original layer and penetrate into a
neighboring one. The dramatic increase in mass transport in the direction of
$\widehat{\bm{n}}$ in combination with nearly perfect nematic order prompted
us to refer to liquid crystals in the RN phase as ``supernematics''
\cite{mazza10}.

\begin{figure}
\psfrag{xx}[c]{$\Delta\overline{r}_{\parallel}(\tau)/\kappa$}
\psfrag{yy}[c]{$\Delta\overline{r}_{\perp}(\tau)$}
\begin{center}
\epsfig{file=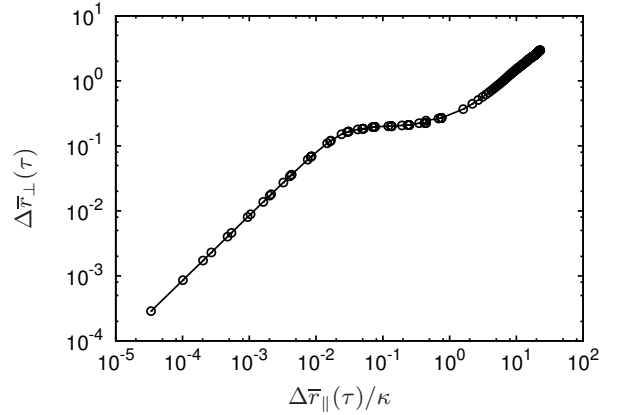,width=0.95\linewidth}
\end{center}
\caption{\small Root MSD $\Delta\overline{r}_{\perp}\left(\tau\right)$ for the
  average displacement of molecules in a direction perpendicular to their long
  axes as a function of the root MSD
  $\Delta\overline{r}_{\parallel}\left(\tau\right)$ for the average
  displacement of molecules in a direction parallel to these axes in units of
  the aspect ratio $\kappa$. Data correspond to the RN phase at $T=4.0$ and
  $P=2.6$. Solid line is a fit to guide the eye.}\label{fig13}
\end{figure}

More detailed insight into the dynamics of mass transport in the RN phase can
be gained by also considering the MSD for molecular displacements in the
direction {\em orthogonal} to the molecules' long axes~\cite{lowoen99} to
which we refer as $\left\langle\Delta
  r_{\perp}^2\left(\tau\right)\right\rangle_t$ by analogy with
Eq.~(\ref{eq:msd}) where, of course, we replace $\bm{r}_i^{\parallel}$ by its
counterpart
$\bm{r}_i^{\perp}\equiv\bm{r}_i-(\widehat{\bm{u}}_i\cdot\bm{r}_i)\widehat{\bm{u}}_i$.
Plots of the root MSD's for perpendicular versus parallel molecular
displacements
$\Delta\overline{r}_{\perp,\parallel}\equiv\sqrt{\left\langle\Delta
    r_{\perp,\parallel}^2\left(\tau\right)\right\rangle_t}$ in
Fig.~\ref{fig13} reveal that dynamically distinct regimes exist. In the
ballistic regime the plot in Fig.~\ref{fig13} increases linearly with slope
$1$ which can be rationalized as follows. Over the relatively short times
characteristic of the ballistic regime molecules move freely in space. Hence,
their self-diffusion is largely controlled by the free volume surrounding a
molecule. In the RN phase the shape of the free volume should be closely
related to the shape of an individual spherocylinder. Therefore, rescaling
$\Delta\overline{r}_{\parallel}\left(\tau\right)$ by the aspect ratio $\kappa$
and expressing $\Delta\overline{r}_{\perp}\left(\tau\right)$ in units of
$\sigma$ one anticipates the plot in Fig.~\ref{fig13} to increase linearly
with a slope of $1$ in the ballistic regime where both
$\Delta\overline{r}_{\perp}\left(\tau\right)$ and
$\Delta\overline{r}_{\parallel}\left(\tau\right)$ are small compared with
$\sigma$. If this condition does not hold any longer intermolecular
interactions begin to matter and self-diffusion is no longer ballistic. The
plot in Fig.~\ref{fig13} shows that during the initial stage of non-ballistic
self-diffusion the plot of $\Delta\overline{r}_{\perp}\left(\tau\right)$
versus $\Delta\overline{r}_{\parallel}\left(\tau\right)$ levels off and
reaches a plateau. During this stage molecules diffuse in string-like
conformations which remain intact, that is the group of molecules forming a
string-like conformation stays together as an entity. However, if both
$\Delta\overline{r}_{\perp}\left(\tau\right)$ and
$\Delta\overline{r}_{\parallel}\left(\tau\right)$ are sufficiently larger than
$\sigma$ the plot in Fig.~\ref{fig13} increases again linearly but with a
somewhat smaller slope of about $0.88$ compared with the initial ballistic
regime. Because in this second linear regime
$\Delta\overline{r}_{\perp}\left(\tau\right)$ is of the order of several
$\sigma$, diffusive motion must involve the exchange of individual molecular
units between neighboring supramolecular ``strings'' which illustrates the
complexity of the dynamics in the supernematic RN phase.

\section{Summary and conclusions}\label{sec:sumcon}
In this paper we elucidate the connection between the formation of string-like
conformations of elongated molecules in the RN liquid-crystalline phase and
enhanced self-diffusivity observed earlier by us \cite{mazza10} in these
phases. We employ a combination of MC and MD simulations performed in an
isothermal-isobaric and in the microcanonical ensemble, respectively. Under
conditions of confinement a sequence of mesophases is observed. These involve
the more conventional I and N phases and, under suitable thermodynamic
conditions, a smA phase. At higher pressures (densities) the more exotic RN
phase may form which is a nematic phase with surprisingly large nematic order
and enhanced self-diffusivity in the direction of the nematic director
$\widehat{\bm{n}}$. Quantitatively, we express the extent of mass transport in
terms of the self-diffusion constant $D_{\parallel}$. In the RN phase
$D_{\parallel}$ may exceed that characteristic of the I, N, and smA phases by
up to an order of magnitude.

Our analysis of the underlying structure reveals that the RN phase is
characterized by so-called ``strings'', that is groups of molecules arranged
such that their long axes and their center-of-mass distance vectors point in
the direction of $\widehat{\bm{n}}$ with only very little deviation from a
perfect arrangement. These arrangements can clearly be seen in representative
``snapshots'' of individual configurations. On the contrary, string-like
conformations are absent as far as I, N, or smA phases are concerned. For the
smA phase this may seem surprising at first on account of the formation of
distinct layers in the direction of $\widehat{\bm{n}}$. However, one has to
bear in mind that in each individual layer the order of center-of-mass
arrangements is still rather low which apparently prevents the formation of
strings between adjacent layers. The reader should also note that if one
considers a {\em single} isolated string that remains intact indefinitely one
would anticipate $\left\langle\Delta
  r_{\parallel}^2\left(\tau\right)\right\rangle_{t}\propto\sqrt{\tau}$
characteristic of single-file diffusion \cite{liu09}. In single-file diffusion
the time dependence of the MSD is caused by the fact that particles cannot
pass each other, a situation encountered experimentally for the motion of
molecules in the narrow spaces existing in materials such as zeolites
\cite{kukla96}, single-wall carbon nanotubes \cite{kaneko08}, or colloids
confined to one-dimensional channels \cite{lutz04}.  However, in the present
case one needs to realize that we are confronted with a totally different
physical situation. As the ``snapshots'' of configurations show many strings
exist which are misaligned in the direction in which they diffuse. At
sufficiently long times individual molecules are exchanged between neighboring
supramolecular ``string-like'' units. As a result, ordinary Fickian diffusion
is the consequence in the limit of sufficiently long times.

The interesting question then becomes: why is this diffusive motion enhanced?
In fact, based upon plots of the equipotential surface one would assume that
even in a string-like conformation molecules would ``feel'' most comfortable
if individual members in both strings are arranged such that neighboring
strings are perfectly aligned because of the energetically favored side-side
arrangement of pairs of molecules. If this were the case one would expect
self-diffusion in the direction of the string-like conformations to actually
be slowed down rather than enhanced on account of restoring forces acting in
the direction of $\widehat{\bm{n}}$ that pull molecules back to relative
arrangements corresponding to locations of potential minima. However, this
rationale ignores the fact that RN phases form at relatively high pressures at
which individual strings are ``squeezed'' together in the direction
perpendicular to the plane of the solid substrates. That strings are actually
squeezed in that direction is reflected by the local density which shows that
in the RN phase one additional layer of molecules can be accommodated compared
with either N or smA phases at lower pressures. In other words, molecules in
neighboring strings are pushed towards each other. This happens to an extent
such that the attractive interactions between any pair of molecules pertaining
to different strings (and therefore the restoring forces in the direction of
$\widehat{\bm{n}}$) are overcome at least partly. In other words, the average distance 
$r^{\perp}$ between the molecules of a pair decreases with $P_{\parallel}$ such that the pair slowly
``climbs up'' the repulsive part of the fluid-fluid interaction potential. Thereby, the net attraction
between the two molecules is reduced as revealed by the shift of the taller maximum of 
$\mathcal{P}(r^{\perp})$ plotted in Fig.~\ref{fig9}. As a result diffusion in
string-like conformations is enhanced rather than diminished.

This mechanism is similar to one proposed for mass transport in zeolites. For example, 
based upon experiments using quasielastic neutron scattering and corresponding MD
simulations, Borah {\em et al.} could demonstrate that in NaY zeolites severe confinement 
causes self-diffusivity to become maximum under geometrically favorable conditions \cite{borah10}.
This effect known as ``levitation'' is most pronounced if the narrowest void in the zeolite
is comparable to the optimum distance between a diffusing molecule and the pore wall which is,
of course, determined by the underlying fluid-wall interaction potential. If the width of the void
is either too small or too large, self-diffusivity is lower than under the optimum confinement conditions.
Levitation is quite regularly observed in zeolites of various kinds 
(see Ref.~\citealp{borah10} and references therein).

At this point one may ask: Is there any experimental evidence supporting the
enhanced self-diffusion that we observe in our simulations? Though the
experimental data so far communicated in the literature do not yet provide a
rigorous confirmation of our theoretical predictions, some of these data are
in reasonable qualitative agreement with our simulations. For example,
extrapolating longitudinal relaxation rates reported in
Ref.~\citealp{miyajima84} from the RN to the N phase yields values markedly
below those in the N phase. If referred to the same $T$ these relaxation rates
correspond to correlation times that are notably shorter in the RN compared
with the N phase. However, this general interpretation of NMR data remains
speculative as long as translational diffusion has not definitely be
identified as the process governing the observed relaxation.

Direct evidence for the relation between translational diffusion and NMR data
can be provided by the pulsed field gradient NMR (PFG NMR) technique
\cite{valiullin09} which records molecular displacements typically over a
$\mu$m range. For example, PFG NMR has been applied to directly assess the
diffusion tensor upon entering the N phase \cite{dvinskikh01}. In these
studies diffusion in the direction of the molecules' long axes was found to
increase with increasing nematic order. By the same technique the diffusivity
of n-alkanes in nanochannels was found to increase with increasing
orientational order \cite{valiullin06}. These findings are in line with our
data where enhanced molecular ordering is accompanied by increasing
diffusivities in the direction of $\widehat{\bm{n}}$. The powerful combination
of PFG NMR with magic angle spinning has recently enabled a notable increase
in both observation times and gradient pulse intensities \cite{maas96}. As a
consequence PFG NMR diffusion measurements became possible beyond the limits
of measurability existing so far. This concerns in particular the first
diffusion measurements with liquid crystals confined to nanopores
\cite{romanova09}. To stimulate a direct experimental verification of our
present predictions using these novel techniques is the primary purpose of
this study.

\begin{acknowledgments}
  We thank Professor J.~K\"arger (Universit\"at Leipzig) for his hospitality
  and fruitful discussions. We also grateful to Lorenzo Guiducci
  (Max-Planck-Institut f\"ur Kolloid- und Grenzfl\"achenforschung
  Potsdam-Golm) for bringing Ref.~\citealp{vollrath01} to our attention and
  thank the International Graduate Research Training Group 1524
  ``Self-assembled soft-matter nanostructures at interfaces'' for financial
  support.
\end{acknowledgments}

\appendix
\section{Theoretical background}\label{sec:theoback}
From a statistical physical perspective the partition function $\mathcal{Q}$
of the canonical ensemble is a key quantity as far as equilibrium systems are
concerned. More specifically, it can be demonstrated that with a proper choice of phase space variables,
$\mathcal{Q}$ for linear molecules can be written as \cite{gruhn97}
\begin{equation}\label{eq:Q}
\mathcal{Q}=
\frac{1}{\Lambda_{\mathrm{th}}^{5N}}
\left(
\frac{\mathcal{I}}{m}
\right)^N
\mathcal{Z}
\end{equation}
where $\Lambda_{\mathrm{th}}\equiv h/\sqrt{2\pi mk_{\mathrm{B}}T}$ is the thermal de Broglie wave length,
\begin{equation}\label{eq:Z}
\mathcal{Z}\equiv
\frac{1}{2^NN!}
\int\int\exp\left[-\beta U(\bm{R},\widehat{\bm{U}})\right]\,\rmd\bm{R}\,\rmd\widehat{\bm{U}}
\end{equation}
is the configuration integral of a system of $N$ indistinguishable molecules,
and $\beta\equiv1/k_{\mathrm{B}}T$ ($k_{\mathrm{B}}$ Boltzmann's constant, $T$
temperature).  The factor $1/2^N$ in Eq.~(\ref{eq:Z}) corrects for double counting equivalent
configurations. These arise because of the head-tail symmetry of the
molecules, that is the equivalence of orientations specified by
$\widehat{\bm{u}}_i$ and $-\widehat{\bm{u}}_i$. For the isothermal-isobaric ensemble employed in
the MC simulations of this work
\begin{equation}\label{eq:Delta}
\Delta\equiv
\sum\limits_{(s_{\mathrm{x}},s_{\mathrm{y}})}
\exp\left(-\beta P_{\parallel}As_{\mathrm{z}0}\right)\mathcal{Q}
\end{equation}
is the relevant partition function \cite{greschek10}. It is then straightforward to show that
the Gibbsian potential \cite{greschek10}
\begin{equation}\label{eq:gibbs}
\mathcal{G}=-k_{\mathrm{B}}T\ln\Delta
\end{equation}
is the relevant thermodynamic potential such that thermodynamic equilibrium is
attained if $\mathcal{G}$ is minimum at any fixed values of $N$,
$P_{\parallel}$, $T$, and $s_{\mathrm{z}0}$. However, it should be noted that
as far as static properties are concerned the specific choice of $m$ or
$\mathcal{I}$ does not matter \cite{greschek10}

However, as far as the {\em temporal} evolution of the system is concerned the
specific choice of both $m$ and $\mathcal{I}$ does matter. To see this let us
derive the equations of motion that govern the temporal evolution in a system
of linear molecules. Starting from Hamilton's equation of motion the
translation of the molecular centers of mass is described by
\begin{subequations}\label{eq:newton}
\begin{eqnarray}
\dot{\bm{r}}_i&=&\bm{v}_i\label{eq:newton1}\\
\dot{\bm{v}}_i&=&\frac{\bm{f}_i}{m},\quad i=1,\ldots,N\label{eq:newton2}
\end{eqnarray}
\end{subequations} 
where $\bm{v}_i$ and $\bm{f}_i=-\nabla_{\bm{r}_i}U$ are velocity and total
force acting on the center of mass of molecule $i$. Similarly, one may derive
equations of motion for the rotation of a molecule starting again from
Hamilton's equations but this time taking angular velocity $\bm{\omega}_i$ and
$\widehat{\bm{u}}_i$ as the conjugate canonical variables rather than linear
momentum and center-of-mass position as in the derivation of
Eqs.~(\ref{eq:newton}). The resulting expressions are given by
\begin{subequations}\label{eq:newtonrot}
\begin{eqnarray}
\dot{\widehat{\bm{u}}}_i&=&\bm{\omega}\times\widehat{\bm{u}}\label{eq:newtonrot1}\\
\dot{\bm{\omega}}&=&\frac{\bm{\tau}_i}{\mathcal{I}},\quad i=1,\ldots,N\label{eq:newtonrot2}
\end{eqnarray}
\end{subequations}
where $\bm{\mathcal{L}}=\mathcal{I}\bm{\omega}$ [see Eq.~(\ref{eq:hamiltonian})],
\begin{equation}\label{eq:torque}
\bm{\tau}_i\equiv\widehat{\bm{u}}_i\times\bm{g}_i=\widehat{\bm{u}}\times\bm{g}_i^{\perp}
\end{equation}
is the torque acting on each molecule, and, following Fincham
\cite{fincham93}, $\bm{g}_i\equiv-\nabla_{\widehat{\bm{u}}_i}U$ is the
so-called ``gorque''. The gorque may be thought of as a turning ``force'' that
effects changes in molecular orientation. For linear molecules obviously only
the gorque $\bm{g}_i^{\perp}$ {\em perpendicular} to the molecular axis
matters. It is obtained through the projection of $\bm{g}_i$ onto
$\widehat{\bm{u}}_i$ via
\begin{equation}\label{eq:gperp}
\bm{g}_i^{\perp}=\bm{g}_i-\left(\bm{g}_i\cdot\widehat{\bm{u}}_i\right)\widehat{\bm{u}}_i
\end{equation} 
Moreover, in the case of linear molecules the angular velocity must always be
perpendicular to the molecular axis, that is
$\bm{\omega}_i\cdot\widehat{\bm{u}}_i=0$. This makes it convenient to replace
the angular by the orientational velocity $\bm{w}_i$ such that
Eqs.~(\ref{eq:newtonrot}) can be replaced by the equivalent but, from a
numerical perspective, much more convenient equations
\begin{subequations}\label{eq:newtonrotfinal}
\begin{eqnarray}
\dot{\widehat{\bm{u}}}_i&=&\bm{w}_i\\
\dot{\bm{w}}_i&=&\frac{\bm{g}^{\perp}_i}{\mathcal{I}}-w^2\widehat{\bm{u}}_i,\quad i=1,\ldots,N\label{eq:newtonrotfinal2}
\end{eqnarray}
\end{subequations}
The derivation of Eqs.~(\ref{eq:newtonrotfinal}) requires a little bit of extra but straightforward vector algebra \cite{fincham93}. Finally, Eqs.~(\ref{eq:newton}) and (\ref{eq:newtonrotfinal}) show that $m$ and $\mathcal{I}$ determine the timescale of translational and rotational motion, respectively.
\bibliographystyle{apsrev4-1}
\bibliography{manuscript-full}
\end{document}